\documentclass[12pt, titlepage]{bbook}

\usepackage{epsfig}
\usepackage{graphicx,color}
\usepackage{amsmath,amsfonts,mathrsfs}

\newcommand{\term}[1]{\emph{#1}}

\newcommand{\pT}{\ensuremath{p_\bot}}
\newcommand{\T}{\ensuremath{_\bot}}

\renewcommand{\exp}[1]{\ensuremath{{\rm exp} \left ( #1 \right )}}
\renewcommand{\ln} [1]{\ensuremath{{\rm ln } \left ( #1 \right )}}

\renewcommand{\cos}[1]{\ensuremath{{\rm cos} \left ( #1 \right )}}
\renewcommand{\sin}[1]{\ensuremath{{\rm sin} \left ( #1 \right )}}
\renewcommand{\tan}[1]{\ensuremath{{\rm tan} \left ( #1 \right )}}
\newcommand{\taq}[1]{\ensuremath{{\rm tan}^2 \left ( #1 \right )}}
\newcommand{\siq}[1]{\ensuremath{{\rm sin}^2 \left ( #1 \right )}}
\newcommand{\coq}[1]{\ensuremath{{\rm cos}^2 \left ( #1 \right )}}
\newcommand{\acos}[1]{\ensuremath{{\rm arccos} \left ( #1 \right )}}
\newcommand{\asin}[1]{\ensuremath{{\rm arcsin} \left ( #1 \right )}}
\newcommand{\atan}[1]{\ensuremath{{\rm arctan} \left ( #1 \right )}}
\newcommand{\sgn}[1]{\ensuremath{{\rm sgn} \left ( #1 \right )}}
\newcommand{\dd}{\ensuremath{{\rm d}}}

\newcommand{\Def}{\ensuremath{\stackrel{\mathrm{def}}{=}}}

\newcommand{\Order}[1]{\ensuremath{\mathscr{O} \left ( #1 \right )}}

\newcommand{\abs}[1]{\ensuremath{| #1 |}}

\newcommand{\ri}[1]{\ensuremath{_{\rm #1}}}

\begin{document}

\frontmatter

\begin{titlepage}\begin{center}

\textsc{{\large Johann Wolfgang Goethe-Universit\"at\\
                Frankfurt am Main}\\
                Fachbereich 13, Institut f\"ur Theoretische Physik}

\medskip

\&

\medskip

\textsc{{\large Universitetet i Bergen}\\
                Institutt for fysikk og teknologi}

\vspace{2cm}

\includegraphics[height=6cm]{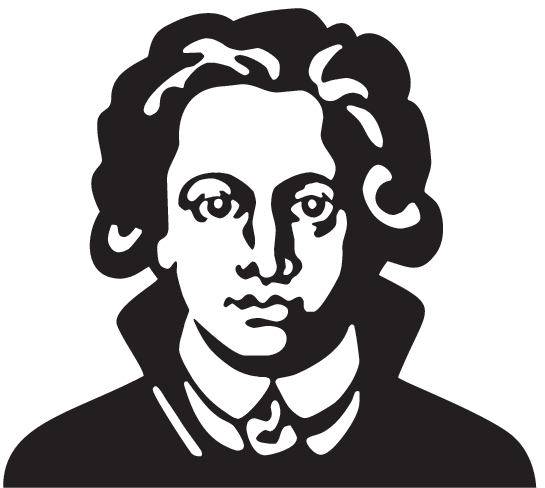}%
\includegraphics[height=6cm]{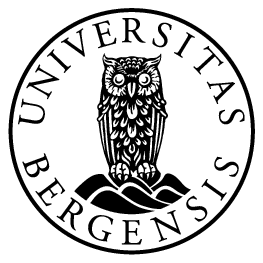}
\vspace{2.5cm}

{\bf \Huge Mach cones in heavy ion collisions}

\bigskip

{\Large Bj\o rn B\"auchle}

\bigskip

{\large June 2007}

\bigskip

Diploma thesis

\bigskip

Advisor: \textbf{Horst St\"ocker} \& \textbf{L\'aszl\'o P\'al Csernai}
\end{center}

\end{titlepage}

\vspace*{8cm}

\begin{center}
{\it F\"ur Hannah}
\end{center}

\cleardoublepage

\tableofcontents
\listoffigures

\mainmatter
\chapter{Introduction}

\section{What are we made of?}

Where we come from, where we are going to and what we are made of are questions
that have been asked by humans at all times. In former days, the only answers
that could be given were speculations by philosophers or claims by priests.
Even nowadays, we cannot answer any of these three questions in a satisfiable
way. But we have shifted the emphasis of them a little.

The quest for our origin is nowadays shifted to evolution and its means, and /
or finally to the big bang that astrophysicists believe has happened about
13.7 billion years ago. No-one can answer the question where that big bang
came from.

Defining our destiny is something that implies the need for forecasts; as our
understanding for the other two questions grows, we gain more and more
knowledge on how to forecast things. Still, a precise forecast on the world's
state tomorrow will never be possible.

The third question remains: What are we made of? We can neither give a final
answer here. But we have come an enormously long way from the believes of the
old greeks, that everything consists of water (Thales) or maybe of the four
elements fire, air, water and earth. In the 19th century, chemists found parts
of the matter they considered indivisible, and they classified about 50 so
called \term{chemical elements} that are built up from those indivisibles
--- the \term{atoms}. Nowadays we know 91 natural chemical elements and have
synthesized several more; their number is about 114. We also found out that
there exist differences between the atoms of a single element; their masses
vary. Atoms of different mass belonging to the same chemical element are
called \term{isotopes}. We know about 2.500 different isotopes.

The sheer variety of classes of atoms (and the fact that we can synthesize
more) indicates that these are not as indivisible as chemists thought in the
first place. Indeed, already in 1909 Lord Earnest Rutherford of Nelson found
that atoms --- he experimented with gold --- consist of a very heavy, positively
charged core, called the \term{nucleus}, and a very light shell, which is
negatively charged. It was soon identified that the shell consisted of the same
particles that make up for electricity, the \term{electrons}. The nuclei, on
the other hand, have been found to consist of \term{protons} and
\term{neutrons}, the so-called \term{nucleons}. Roughly spoken, the number of
protons in a nucleus determines the chemical element, whereas the number of
neutrons distinguishes different isotopes. By those discoveries the variety of
2.500 different indivisible particles has been reduced to 3\footnote{For
completion, it should be said that the electrons make up for the chemical
properties of an element, but \emph{their} number is determined by the number
of protons in the nucleus.}.

When trying to study electrons, protons and neutrons, further particles have
been discovered. Besides particles that resemble light --- the so-called
\term{photons} --- and similar particles, the new particles can be put into
two classes: Those that interact with nuclear matter as nuclear matter
itself does, i.e.\ with the so-called \term{strong force}, and those that
don't. The first are called \term{hadrons}, the latter are
\term{leptons}\footnote{The terms have greek origin and mean ``strong
particles'' (hadrons) and ``light particles'' (leptons), referring to the
strong force and low masses, respectively.}. In the \term{standard model of
particle physics}, which is the state-of-the-art-theory of our knowledge
about matter, the six known types of leptons (called \term{electron},
\term{muon}, \term{tauon}, \term{electron-neutrino}, \term{muon-neutrino}
and \term{tau-neutrino}) remain fundamental parts of matter. No experimental
results indicate that they have an inner structure.

This is not the case for the hadrons. Hundreds of hadrons have been directly
or indirectly detected, and it turned out that they are not fundamental.
Instead, new degrees of freedom have been postulated and experimentally
observed, so-called \term{quarks} and \term{gluons}. In accordance to the
number of leptons, we know six quarks, called \term{up}, \term{down},
\term{charm}, \term{strange}, \term{top} and \term{bottom}, from which only up
and down form neutrons and protons. Gluons are to the strong interaction what
photons are to electrodynamics, the theory that covers electricity, magnetism
and light.

Quarks and gluons are, as leptons and photons, considered fundamental parts of
matter. Their observation, though, is not very easy. Due to the structure of
the strong force and the underlying theory, the \term{Quantum Chromo Dynamics}
(QCD), they can never be seen alone, but only in pairs (one quark and an
antiquark) or in triplets (three quarks or three antiquarks), or any
combination thereof. This phenomenon is called \term{confinement}. When scatterings
between quarks happen at very high center of mass-energies, this confinement
ceases to exist. This is called \term{asymptotic freedom}. It is subject to
experimental research to create \term{deconfined} matter, i.e.\ matter that is
asymptotically free. Since this matter consists of free quarks and gluons, it
is called \term{Quark Gluon Plasma} (QGP).

\section{Studying Quark Matter}

The purpose of the largest experiments on earth, such as the Relativistic Heavy Ion
Collider (RHIC) at Brookhaven National Laboratory (BNL), the Tevatron at Fermilab
and --- soon --- the Large Hadron Collider (LHC) of the European Council for Nuclear
Research (CERN), is to probe the smallest parts of matter we know. At these
facilities, protons (Tevatron and LHC), gold- (RHIC) and lead nuclei (LHC) are
accelerated to velocities very close to the speed of light and collided with
each other. In proton-proton-collisions as the smallest possible system of
colliding stable hadrons, the fundamental forces and particles can be studied in
a very clean way, whereas in the bigger systems (gold on gold and lead on
lead) collective properties of the matter can be studied.

The latter is the kind of physics that is addressed in this thesis. In systems
with 396 or 416 initial nucleons (gold and lead, respectively), it can e.g.\
be studied how the single particles react in connection with and surrounded by
many other particles. Quantum Chromo Dynamics (QCD), the gauge theory for strong
interactions, predicts free quarks and gluons in the infinite high temperature
limit. It is hoped that this ``deconfinement'' can be reached in collider
experiments. Indeed it is claimed to have been seen in experiments at the
Super Proton Synchrotron (SPS) at CERN \cite{else:PR_SPS_QGP} and at RHIC.

\begin{figure}\begin{center}
\includegraphics[width=.75\textwidth]{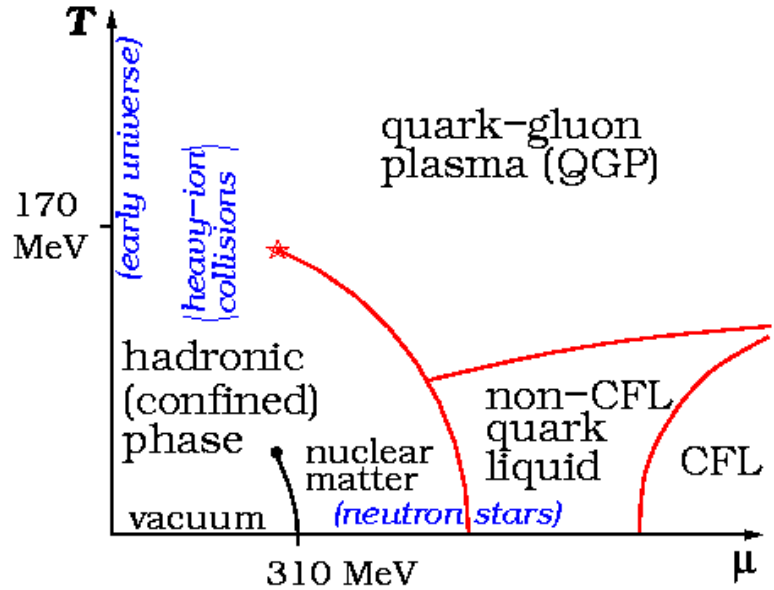}
\caption[Phase diagram of QCD]{The phase diagram of QCD. It shows the
different expected phases. At $\mu = 310$~MeV and $T = 0$ lies ground state
matter. Note that the position of phase transitions is not known
exactly. From~\cite{else:wiki_PD_QGP}}
\label{fig:phasediagram}
\end{center}\end{figure}

When studying heavy ion reactions, a major field of interest is the
so-called \term{phase diagram} of QCD. Like in everyday physics, such a
diagram shows in what state the matter is at given conditions. Unlike in
everyday physics, the conditions are not given in pressure and temperature,
but in (baryo-)chemical potential $\mu_B$ and temperature $T$ (see figure
\ref{fig:phasediagram}). Depending on the \term{center of mass-energy}
$\sqrt{s_{NN}}$ of the collision and on the system size (determined by the
nuclei involved) one can probe different regimes of the diagram. With very
high energies and big nuclei the matter is heated very much (high $T$), but
has few baryons (low $\mu_B$). Such matter is created at RHIC and LHC. Here,
a second order phase transition is believed to occur at a temperature $T_C
\approx 170$~MeV \cite{Aoki:2006br}.

This phase-transition leads from a gas of normal, confined hadrons, the
\term{Hadron Gas} (HG), to deconfined Quarks and Gluons, the
\term{Quark-Gluon-Plasma} (QGP). This phase seems to behave like a liquid,
according to claims made by the experiments at RHIC \cite{else:PR_RHIC_QGP}.
This is surprising, because it has previously been thought of as a perfect
gas.

The theoretical problem one has when dealing with nuclear matter is that QCD
cannot be solved perturbatively at small energy-momentum transfers $Q^2$.
This is due to the at low $Q^2$ large coupling constant. Unlike the
electromagnetic coupling the strong coupling varies a lot and actually falls
with increasing $Q^2$, which is the reason for the theory being
``asymptotically free''. For lower temperatures (and therefore lower $Q^2$),
nuclear systems cannot be calculated from first principles (namely QCD), but
with effective models. Some of these are described briefly in
chapter~\ref{chap:collectivemodels}. Also, the concept of temperature itself
introduces the need for many particles, and there is no kind of interaction
in which many-particle systems can be described exactly.  This is another
reason for the need of effective models, even above the threshold for
deconfinement.

Of the existing models for dynamical descriptions of heavy-ion reactions
(see chapter~\ref{chap:collectivemodels}), hydrodynamics is very popular. In
this approach, the matter is described as a fluid. Hence, collective effects
can be studied easily. For example, collective flow is an observable
intimately connected to hydrodynamic behaviour (for more detail, see
section~\ref{sec:hydro}).

\section{Supersonic acoustic sources}

In all fluids, there exists a mode for propagating weak, linear perturbations.
Those perturbations are exactly what excites the human eardrums and causes us
to hear --- \emph{sound}. The speed with which these perturbations move is the
largest speed by that any mechanical stimulus can travel through a given body
(which does not necessarily have to be a liquid) and depends on the material
the body consists of. It might also depend on the wavelength of the
perturbation, this phenomenon is known as dissipation and does not exist in
perfect liquids.

The speed of sound $c_S$ can be calculated from the Equation of State (EoS,
see section \ref{par:hydro_equations}, page \pageref{par:hydro_equations})
to be the partial derivative of the pressure $p$ with respect to the energy
density $e$
\begin{equation}\label{eq:speed_of_sound}
c_S^2 = \frac{\partial p}{\partial e}\quad.
\end{equation}

When an acoustic source moves through a medium, the audible sound changes. A
resting observer in front of the source will hear a higher frequency than is
actually emitted, if the source passed her, she will hear a lower frequency.
This phenomenon is known as the \term{Doppler-effect}. It can be easily
understood if one considers the sound waves emitted by the source as being
compressed in forward direction and elongated in backward direction.

\begin{figure} \begin{center}
\includegraphics[width=.75\textwidth]{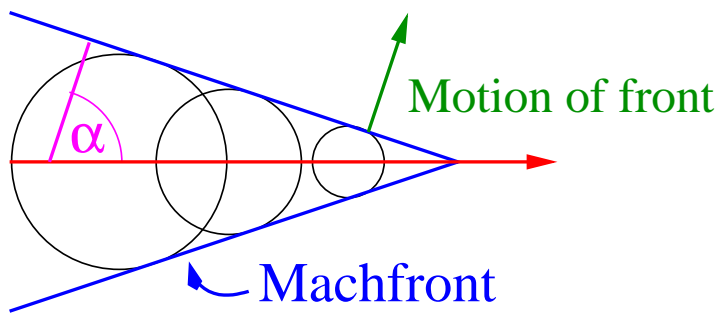} \end{center}
\caption[Sketch on mach cones]{A schematic view on how a mach cone develops
(in two dimensions): The elementary sound waves all add up at one line. The
angle $\alpha$ is given by $\cos{\alpha} = c_S / v$, where $v > c_S$ is the
velocity of the source.} \label{fig:machcone_explanation} \end{figure}

If the speed of the source is the same as the sound velocity, all sound
waves ever emitted by it will reach the observer at the same instant ---
together with the source itself. One cannot hear no frequency anymore, but
only a \term{(super)sonic boom}. In the supersonic regime, when the source
is moving faster than the sound it emits, the sound arrives after the source
has passed. The waves form a cone --- the
\term{mach cone}, named after Ernst Mach, an Austrian physicist who lived
from 1838 until 1916. The cone develops because there is a straight surface
perpendicular to every elementary sound wave emitted at any instant, see
figure \ref{fig:machcone_explanation}. 

Unlike normal sound perturbations, the amplitude of the mach cone does not
decrease with the square of the distance to the exciter, since it is not a
point, but only with the distance to the first power, since the exciter
forms a straight line (this is analogous to the electrical field of a
pointlike charge and a infinitely long, homogeneously charged wire). 

Measuring the angle of a mach cone will give insight on the speed of sound
or the speed of the particle, if the respectively other is known. Therefore
it might help to falsify an assumed Equation of State for the medium
considered.

Whereas in everyday experience, the velocity of the medium, say, air, is
small in comparison to the speed of sound and its fluctuations therefore are
small as well, this is not the case in general. In nuclear matter,
collective motion may be as fast as and even faster than the speed of sound
and vary a lot along the trajectory of the sound source. To study how much
this affects the resulting shape of a mach cone is the goal of this thesis.

In our special case, we consider a high energy jet, as might result in a
partonic collision in the very early stages of a heavy ion reaction, and
examine the sound waves it emits and their way through a realistic medium
that moves with a speed comparable to the speed of sound itself (something
that will never be seen in everyday physics).

\chapter{Models for heavy ion collisions}\label{chap:collectivemodels}

\section{Thermal model}\label{sec:thermal}

Since the transverse momentum spectra of particles in heavy ion collisions are
pretty similar to what is predicted by thermodynamical assumptions, they are
sometimes described as coming from a source that is globally equilibrated
\cite{Becattini:2001fg,Braun-Munzinger:1994xr,Braun-Munzinger:1995bp,
Cleymans:1999st,Kisiel:2005hn,Torrieri:2004zz}.

The model starts with the phase space density, which is given as
\begin{equation}\label{eq:f_eq}
f(x, p) = \frac{\dd N}{\dd \Gamma} = \frac{\dd N}{\dd^3 x\, \dd^3 p} =
\frac{g}{(2 \pi)^3}\frac{1}{\exp{\frac{E-\mu}{T}} + \alpha}\quad.
\end{equation}

Here, $E = \sqrt{\vec p^2 + m^2}$ is the energy, which will in the more
general case of a moving source (moving with the four-velocity $u^\mu$) be
replaced by the scalar product $p_\mu u^\mu$. $g$ and $\mu$ denote the
degeneracy factor and chemical potential for the particle species considered,
and $T$ is the temperature. The latter is assumed to be the same for all
species. $\alpha$, finally, distinguishes the different spin statistics: For
bosons (integral spins), it is $\alpha\ri{BE} = -1$, which results in the
Bose-Einstein-distribution, whereas for fermions (half-integer spin) it is
$\alpha\ri{FD} = +1$, which leads to the Fermi-Dirac-distribution. For high
energies and high temperatures both distributions become equal, $\alpha$ can
be neglected, and one obtains the Maxwell-Boltzmann-distribution.

A thermal model is by far the most macroscopic approach to heavy ion
collisions. It can not account for local inhomogeneity and its applicability
to a system exploding immediately and with high velocities is highly
questionable.

\section{Transport model}\label{sec:transport}

An approach that accounts not only for local deviations from an assumed
global symmetry, but really tries to model each particle and its trajectory
through the space time, is the transport approach
\cite{Bass:1998ca,Bleicher:1999xi,Bratkovskaya:2004kv}. The evolution of the
system follows the relativistic transport (or \emph{Boltzmann}-) equation
\begin{equation}\label{eq:transport} p^\mu \partial_\mu f(x, \, p) = {\rm
St}\left \{f \right \} \end{equation}

Here, ${\rm St} \left \{f \right \}$ is called the collision term, which
contains information on cross-sections and acts as source term for the
density function.

A set of solutions to that formula is the equilibrium phase space density
used for thermal and hydrodynamical models \eqref{eq:f_eq}.

Transport models assume point like, classical particles whose mean free path
is very large. The description of three-particle-collisions is very
complicated.

\section{Lattice QCD}\label{sec:lattice}

An approach to solve the equations of QCD directly is to perform lattice
calculations. Here, space and time are descretised (therefore ``lattice'').
The properties of infinitely vast matter in equilibrium, and therefore the
Equation of State, can be
studied \cite{Boyd:1996bx,Karsch:2000kv,Allton:2002zi,Fodor:2001pe,Fodor:2001au}.
It does not provide a dynamical description, but can give exact input to
hydro- and thermodynamical models.

\section{Hydrodynamical model}\label{sec:hydro}

A model widely used to describe heavy-ion reactions is fluid- or
hydrodynamics. It has been predicted as a key mechanism for the creation of
hot and dense matter very early \cite{Hofmann:1975by,Hofmann:1976dy}. Using
this approach one assumes that the matter described is in local thermal
equilibrium or at least in a state showing only small deviations from that.
This assumption is not obviously met in a heavy ion reaction. Indeed, seen
from any frame, the particle distribution functions of projectile and target
are very different and far from being equilibrated.  In the progress of the
collision, though, a locally equilibrated system may be formed. The part of
the reaction before equilibration can therefore not be described by usual
(1-fluid-) hydrodynamics.

One approach is to introduce several distinct fluids that each hold a
equilibrated subsystem, e.g.\ one fluid for the target, one for the projectile
and optionally a third for the evolving fireball. The whole system here is not
equilibrated, but hydrodynamical evolution of each component is possible.
Interaction between the components is implemented by considering the other
components as source terms in the equations.

This approach allows for modelling of transparent nuclei, i.e.\ of the fact
that at low energies the nuclei penetrate through each other.

In so-called \term{one-fluid hydrodynamics} the beginning of the reaction
cannot be modelled. Therefore, an additional model has to be applied for the
creation of the first equilibrated state, the so-called \term{initial
state}. This can in principle be any kind of non-equilibrium model. The
creation of the first equilibrated state can take different times, depending
on the mechanism with which it is reached. Times of the order of $\tau_0
\approx 1$~fm are reasonable.

When the initial state is defined, hydrodynamics start to work. Depending on
number and kind of the assumed symmetries in the initial state the fluid
development may be solvable analytically or only numerically. Within this
stage of the calculations, one has to assume an equation of state (EoS).
It must also be chosen if the evolution describes a perfect fluid that is
perfectly equilibrated at any point or if small deviations are endorsed. In
the latter case, several additional parameters like the heat conductivity,
shear- and bulk-viscosity have to be introduced.

In the case of a heavy ion collision the reaction zone is surrounded by
vacuum, into which the matter will expand.  Therefore, the energy-density and
the particle density will decrease as a function of time. The system gets more
and more dilute, and the assumption of local equilibrium at any point gets
more and more unjustified. A condition has to be defined where the
hydrodynamical evolution is stopped and something else is done. This step is
called \term{freeze-out}. Whatever the condition is --- specific lab- or
proper-time, temperature, energy density, baryon number density ---, it must be
such that each part of the system crosses that condition and can be frozen
out. The area in which freeze-out happens is usually a three dimensional
\term{hyper-surface}, but it might have a finite thickness, which would lead to
a hyper-layer or hyper-volume.

Apart from the choice of the position of the surface, some other things may be
adjusted for the freeze-out process. This starts with the kind of matter in
the final state. Here, massive hadrons, massive quarks (without hadronisation
so far), even massless quarks or hadrons may be assumed. Quantum effects in
the distribution function may or may not be taken into account.

In any case, freeze out should conserve some quantities. Besides
energy-momentum- and baryon number-conservation care has to be taken that
entropy does not decrease. It is an additional constraint to require even the
gross baryon number (number of baryons plus number of antibaryons) to be equal
across the freeze-out process.

Even for the surface it is not enough to state its position, its thickness
has to be taken care of as well. While the classical, standard approach
considers it to be infinitely small, it may also be that freeze-out happens
at a whole layer, which implies that particles may leave the system that are
close to, but not at the bordering condition.

\bigskip\label{par:hydro_equations}

Starting with the Boltzmann transport equation \eqref{eq:transport} with a
vanishing source term, one can define \term{moments} of the distribution as
\begin{eqnarray}\label{eq:Nmu}
N^\mu(x) & = & \int \frac{\dd^3 p}{p^0} p^\mu f(x, p) \\
\label{eq:Tmunu}
T^{\mu\nu} (x) & = & \int \frac{\dd^3 p}{p^0} p^\mu p^\nu f(x, p)
\end{eqnarray}
which are the baryon number current and the energy-momentum density,
respectively. Since $p^\mu \partial_\mu f = 0$ (see \eqref{eq:transport}),
it follows that
\begin{eqnarray}\label{eq:hydro_contin1}
\partial_\mu N^\mu =& 0\\\label{eq:hydro_contin2}
\partial_\mu T^{\mu\nu} =& 0 & .
\end{eqnarray}

These five equations (there is one for every $\nu$ in
\eqref{eq:hydro_contin2}) face fourteen independent variables whose time
developments have to be found (four independent components of $N^\mu$ and 10
independent components of the symmetric four by four-tensor $T^{\mu\nu}$).
By assuming perfect local thermal equilibrium 8 independent components of
$T^{\mu\nu}$ can be eliminated. Then, one is left with 6 parameters and only
needs one more equation --- the equation of state (EoS). This usually gives a
relation between pressure, energy density and baryon number density. For a
detailed description on how to decompose $T^{\mu\nu}$ into a ideal and
non-ideal part and the different possibilities for various definitions
please refer to \cite{Rischke:1998fq,Cse92}.

Very simple equations of state consider ideal, ultra-relativistic (i.e.\
massless) gases. Here, the speed of sound is constant $c_S^2 = 1/3$.
The pressure is not dependent on the baryo-chemical potential. This EoS is
considered good in the QGP-regime. For lower temperatures, the simplest case
is a (massive) hadron-resonance-gas. Here, the speed of sound is given by $c_S^2 =
0.15$.

\bigskip

\paragraph{The hydrodynamical model}\label{par:themodel} used in this thesis
is the \term{particle in cell} (PIC)-method which has been developed by
Harlow and Amsden in the early 1960s \cite{Amsden1,Amsden2} and upgraded to
ultra-relativistic energies by Nix and Strottman in the 80s and 90s
\cite{Clare:1986qj,Strottman:1989ln,Amelin:1991kb}. It combines the
advantages of fixed cells (Eulerian grid) and free moving particles
(Lagrangian markers) and has e.g.\ explicit baryon number conservation. An
ideal gas with two flavours is assumed for the Equation of State. It has a
constant speed of sound $c_S = 1/\sqrt{3}$. The initial state has been
developed recently by Magas, Csernai and Strottman and is exhaustively
explained in \cite{Magas:2000jx}. Freeze-out happens at constant time and
goes from massless QGP to massive quarks with a restmass of $m_q \approx
300$~MeV.  Hadronisation is abjured \cite{Zetenyi:pc}. It conserves energy,
baryon number and gross baryon number.

\chapter{Jets and Medium}\label{chap:jets}

In any collisions of the kind $2 \rightarrow 2$ (i.e.\ two incoming particles
produce two outgoing particles), the daughter particles will be, seen from the
center of mass system, back-to-back-correlated. This is a simple consequence
of conservation of momentum. When the momentum transfer $Q^2$ in a collision
becomes sufficiently large, the structures that play a role for scatterings
become small enough not to resolve whole hadrons, but their constituents, the
partons. In other words, at high energies partonic interactions dominate over
hadronic interactions. In this region also \term{perturbative QCD} (pQCD) is
applicable, because the strong coupling constant $\alpha\ri{S}$ is small
enough.

When two partons scatter, their daughter particles usually fragment into
hadrons after a short time, that means they create new $q\bar q$-pairs from
the vacuum, thereby losing energy and form hadrons themselves as well as new
hadrons, that typically go parallel to the leading hadrons. Theses bunches of
hadrons are called jets.

If nothing hinders the jets from being detected, as is the case in
proton-proton collisions, two jets will be seen that are usually exactly
back-to-back-correlated. The center of mass of very high
energetic partonic collisions is usually the same as the center of mass of the
proton-proton collision.

When correlating the azimuthal angles $\varphi$ of the observed hadrons one
can recognize such events by two peaks at $\Delta \varphi = 0$ and $\Delta
\varphi = \pi$. This signal for \term{di-jets} is seen in $pp$-collisions
at high energies.

In heavy ion collisions, it is not so clear whether such a signal can be seen.
The created jets might have to penetrate through the medium that is present in
such reactions, except when the initial parton-parton collision happens at the
surface of the medium and both daughter particles go tangential to the
surface. In all other cases, at least one of the jets has to go through the
medium. Indeed, the most interesting case is a hard collision close to the
surface with one daughter going right out of the collision zone and the other
going the longest possible way through it. Here, studying the so-called
\term{near-side-jet} (the one leaving the system immediately) can give
insight on the parameters of the original hard collision and on the parameters
of the \term{away-side-jet}.

Depending on the properties of the medium the away-side-jet will look
different to the experiment. For example, the cross-section for interactions
between jet and medium might be very small, or the medium dilute enough, so
that the jet goes through the medium almost undisturbed. This is the case at
(comparatively) low energies.

\begin{figure}\begin{center}
\includegraphics[width=.5\textwidth]{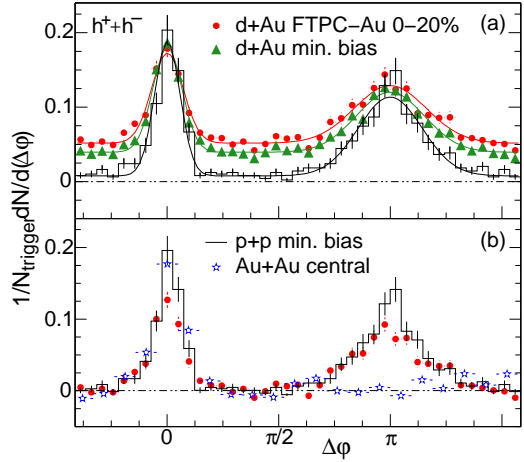}
\caption[Di-Jet-Suppression at $\sqrt{s_{NN}} = 200$~GeV]{Two particle
correlations for p+p, d+Au and Au+Au-collisions at $\sqrt{s_{NN}} =
200$~GeV. Trigger particles are in $4\,{\rm GeV} < \pT{}\ri{Trig} < 6\,{\rm
GeV}$, and associated particles are all those with a momentum bigger than 2
GeV and less than the trigger $2\,{\rm GeV} < \pT{}\ri{assoc.} <
\pT{}\ri{Trig}$.  The curves show a clear suppression of secondary particles
at $\Delta \varphi = \pi$ in gold-gold-collisions when compared to the smaller
systems. From \cite{Adams:2003im}.} \label{fig:star_suppression}
\end{center}\end{figure}

At central RHIC collisions, however, there is no away-side-jet observed with
an energy comparable to the near-side-jet \cite{Adams:2003im}, see figure
\ref{fig:star_suppression}.

Obviously, the away-side-jet loses its energy, which is in return absorbed in
some way by the medium. Different models for this energy loss exist, and the
exact mechanisms are subject to current discussion in the field. A long time
\cite{Baier:1996sk,Baier:1996kr,Zakharov:1996fv,Gyulassy:2000fs,Gyulassy:2000er}
radiative energy loss has been considered the dominant mechanism. In 2003,
Mustafa and Thoma \cite{Mustafa:2003vh} re-considered energy loss by
collisions, as had been predicted long before
\cite{Bjorken:1982tu,Thoma:1990fm,Braaten:1991jj,Braaten:1991we}. Only
recently, Mustafa and Thoma's results gained more attention
\cite{Wicks:2005gt}. Peshier
\cite{Peshier:2006hi,Peshier:2006mp,Peshier:2006az,Peshier:2006ah} has shown
in 2006 that collisional energy loss is indeed very important for the
explanation of the magnitude of transport coefficients etc.

\begin{figure}\begin{center}
\includegraphics[width=.5\textwidth]{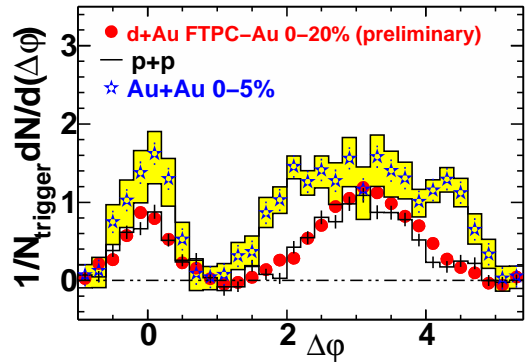}
\caption[Sideward peaks in two-particle correlations]{Two particle
correlations for p+p, d+Au and Au+Au-collisions at $\sqrt{s_{NN}} =
200$~GeV with same threshold for trigger particles as in figure
\ref{fig:star_suppression}, but lower trigger for associated particles
$0.15\,{\rm GeV} < \pT{}\ri{assoc.} < 4\,{\rm GeV}$. The data show sideward
peaks at $\Delta \varphi \approx \pi \pm 1$. From \cite{Wang:2005hp}.}
\label{fig:star_twoparticle}
\end{center}\end{figure}

In any case, secondary particles will have significantly lower momentum than
the near-side-jet, which is a good explanation to the apparent complete
disappearance of the away-side-jet in \cite{Adams:2003im}. When lowering the
threshold for secondary particles, one can see sideward peaks in the
two-particle azimuthal correlations \cite{Adams:2005ph}, see figure
\ref{fig:star_twoparticle}. Such a signal has been predicted as signature
for mach shocks in 2004 \cite{Stoecker:2004qu}.  But also large-angle gluon
radiation \cite{Vitev:2005yg,Polosa:2006hb}, jets deflected by radial flow
and \v cerenkov radiation \cite{Dremin:2005an,Majumder:2006dp} are
consistent with the observed away-side structure.

\begin{figure}[t]\begin{center}
\includegraphics[width=.5\textwidth]{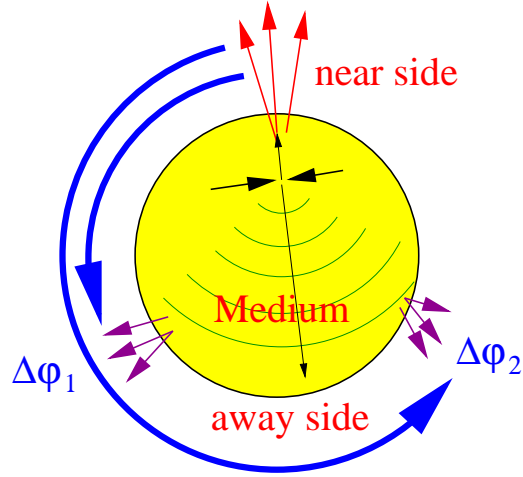}
\caption[Scheme on three-particle correlations]{This sketch shows the angles
used for three-particle correlations. Azimuthal differences from two different
particles to the high-$\pT$-particle are opposed.}
\label{fig:threeparticle_explanation}
\end{center}\end{figure}

Mach cones in nuclear matter have been predicted for cold nuclear matter
\cite{Hofmann:1975by,Stoecker:1979iv,Stoecker:1986ci,Rischke:1990jy,Chapline:1986sw},
fermi liquids \cite{Glassgold:1959XX,Khodel:1980yy} and QGP
\cite{Casalderrey-Solana:2004qm,Casalderrey-Solana:2007km,Antinori:2005tu} and
observed in heavy ion reactions at RHIC
\cite{Adler:2002tq,Molnar:2007wy,Adler:2006sc,Wang:2004kf,Ulery:2006ix,Wang:2006xq,Wang:2006ig,Jacak:2005af,Ajitanand:2005xa}.

Recently, the predictions are supported by measured three-particle
correlation spectra, where two azimuthal correlations are being opposed (see
figure \ref{fig:threeparticle_explanation}). Here, a clear distinction
can be made between an indifferent enhancement of spectra by deflection
etc.\ and mach cones or \v cerenkov radiation that pronounce explicit
directions. Peaks at $(\Delta \phi_1, \Delta \phi_2) = (\pi \pm b, \pi \pm
b)$ (on the bisector) would be present in all scenarios, they correspond to
the fact that a jet typically consists of more than one particle, and a peak
on the bisector merely says that two particles are emitted at the same
angle. Peaks at $(\Delta \phi_1, \Delta \phi_2) = (\pi \pm b, \pi \mp b)$
(note the opposite sign!), on the other hand, show a correlation between
markers on \emph{both sides} of the backward direction.

To distinguish between \v cerenkov radiation and mach cones, one can look at
the \pT-dependence of the angle. In \cite{Dremin:2005an} it is shown that
the angle of the \v cerenkov cone should be increasing very quickly with the
momentum of the associated particles. 

\begin{figure}\begin{center}
\includegraphics[width=.5\textwidth]{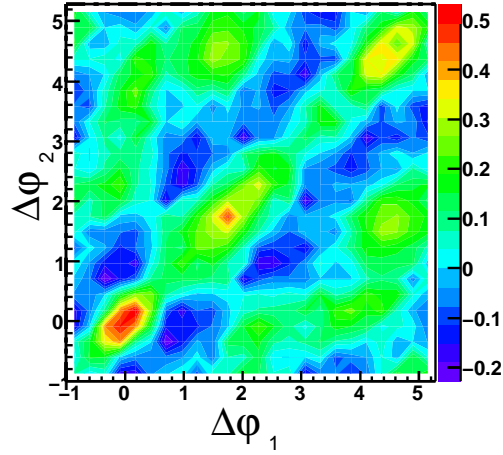}
\caption[Three-particle correlations from STAR]{Three-particle correlations
from STAR. $b$ (see text) is about 1.1 radians $\equiv
63^\circ$ here, which corresponds (in static medium) to a speed of sound of
$c_S^2 \approx 0.21 \pm 0.8$. See text for more details. From
\cite{Ulery:2007zb}.}
\label{fig:star_threeparticle} \end{center}\end{figure}

Preliminary data from STAR (see figure \ref{fig:star_threeparticle}) support
the conical mechanisms (mach cones or \v cerenkov radiation), and deeper
insight leave, according to \cite{Ulery:2007zb}, little doubt on the
dominance of mach cones over \v cerenkov radiation. 

\bigskip

Satarov \cite{Satarov:2005mv} and Chaudhuri \cite{Chaudhuri:2006qk} have
calculated that the jet angle in an expanding medium is also very dependent
on the exact origin of the jet. In short, the angle rises when the jet does
not come from the middle of the medium.

\section{Calculated correlations in static medium}\label{sec:analytical}

It is not trivial to calculate the expected angular distribution even for
static medium. The reason for that is that one has to use two different
coordinate systems, both of which are spherical. One is the laboratory
system, having the beam axis as the pole, the other one having the jet axis
as pole. Describing the cone is very easy in the latter, which will be
denoted $(\alpha, \beta)$ for polar- and azimuthal angle, but the measured
angle will be seen from the outside system $(\vartheta, \varphi)$. Both
systems never are the same, because the jet can --- in the model used ---
never go into the beam direction (see chapter \ref{chap:model}).

A jet propagating through static medium will cause a cone appearing at a
constant longitude, in the system where the jet is going towards the pole.
The angular distribution is here
\begin{equation}\label{eq:spectrum-static-ab} \frac{\dd N}{\sin{\alpha} \dd
\alpha \, \dd \beta} = \delta \left ( \cos{\pi - \alpha} - c\ri{S} \right
)\quad, \end{equation}
or, in other words, the cone is at
\begin{equation}\label{eq:cone-position-ab} {\hat r}_{\alpha\beta} = \left
(\begin{array}{ccc} \sin{\pi - \alpha_0}\cos\beta \\ \sin{\pi -
\alpha_0}\sin\beta \\ \cos{\pi - \alpha_0} \end{array} \right )\quad,
\end{equation}
where the index $\alpha\beta$ defines the coordinate system to have the
$\hat z$-axis in the direction of the pole, which is the direction of the
jet. Here, $\cos{\pi - \alpha_0} = -\cos{\alpha_0}$ is fixed to the speed of
sound (it is the usual mach angle), but $\beta$ runs around the circle:
$\beta \in [0;\,2\pi)$. Each point on this circle is equally weighted with
spectral enhancement, so that $\dd N / \dd \beta = const$. Note that, as long
as one stays consistent within one consideration, the exact choice on the
$\beta = 0$-direction is arbitrary, therefore, at the start, $\cos\beta$ and
$\sin\beta$ are interchangeable. A relative negative sign will occur if a
jet in positive and negative beam direction are considered, because the
direction of rotation of $\beta$ has been changed.

In order to obtain the spectrum $\dd N / \dd \varphi$ as measured in a
heavy ion experiment, one has to rotate the system $\alpha\beta$ to
$\vartheta\varphi$. For a jet at mid-rapidity this is simple. In the
following, the jet will always go in $\hat x$-direction, where the azimuthal
angle is zero (this is equivalent to using the difference angle $\Delta
\varphi$). Then, the cone will be at
\begin{equation}\label{eq:cone-position-mid-pt}
{\hat r}_{\vartheta\varphi} = \left (
\begin{array}{c}
 \cos{\alpha_0} \\
 \sin{\alpha_0} \cos\beta \\
 \sin{\alpha_0} \sin\beta
\end{array} \right )\quad.
\end{equation}
Now, the azimuthal angle $\varphi$ can be read off:
\begin{equation}\label{eq:phi-from-mid}
\tan\varphi = \frac{y}{x} = \tan{\alpha_0} \cos\beta\quad.
\end{equation}
$\dd N / \dd \varphi$ can be expressed as $\dd N / \dd \beta \cdot \dd \beta
/ \dd \varphi$, where the first factor is one (see above). The latter can be
obtained from equation \eqref{eq:phi-from-mid}:
\begin{eqnarray}
\beta &=& -\asin{\frac{\tan\varphi}{\tan{\alpha_0}}}\\
\frac{\dd \beta}{\dd \varphi} &=& \frac{-\,(1 + \taq\varphi)}{\sqrt{\taq{
\alpha_0} - \taq\varphi}}\quad.\label{eq:dbetadphi-mid}
\end{eqnarray}
For the minus-sign in equation \eqref{eq:dbetadphi-mid} refer to the
statement above.

In the case of a jet that is not going to mid-rapidity, the formul\ae{}
become a lot more complicated. To characterize the direction of the jet, we
take the angle $\tau$ between the direction of the jet and mid-rapidity.
Thus, $\tau = \pi / 2 - \vartheta$. The general idea is the same as before;
we rotate ${\hat r}_{\alpha\beta}$ (see equation
\eqref{eq:cone-position-ab}) to ${\hat r}_{\vartheta\varphi}$, solve
for $\beta$ and derive with respect to $\varphi$. 

The rotation now happens with a more complicated matrix; we rotate around
the $\hat y$-axis with an angle of $\tau$:
\begin{eqnarray}
{\hat r}_{\vartheta\varphi} &=& \left (
\begin{array}{ccc}
 \cos{\tau}   & 0 & + \sin{\tau} \\
 0            & 1 & 0\\
 - \sin{\tau} & 0 &   \cos{\tau}
\end{array} \right ) \left ( \begin{array}{c}
 \cos{\alpha_0} \\
 \sin{\alpha_0} \cos\beta \\
 \sin{\alpha_0} \sin\beta
\end{array}\right )\\
&=& \left ( \begin{array}{c}
\cos\tau \cos{\alpha_0} + \sin{\alpha_0} \cos\beta \sin\tau\\
\sin{\alpha_0} \sin\beta\\
\cos\tau \sin{\alpha_0} \cos\beta - \cos{\alpha_0} \sin\tau
\end{array}\right )\quad.
\end{eqnarray}
As before, we read off $\tan{\varphi}$ as the ratio of y- and x-component.
Using $A \cos\phi + B \sin\phi = \sqrt{A^2 + B^2}\cos{\phi - \atan{B / A}}$
and basic geometry one finds 
\begin{eqnarray}\nonumber
\beta& = &\acos{\frac{-\,\cos\tau \tan\varphi}{\tan{\alpha_0}}\sqrt{\frac{1}{1 +
\siq\tau \taq\varphi}}}\\
& + & \atan{\frac{1}{\sin\tau \tan\varphi}} \label{eq:beta_tau}
\end{eqnarray}
and, after some more calculations,
\begin{eqnarray}\nonumber
\frac{\dd \beta}{\dd\varphi} &=&
\frac{1+\taq\varphi}{1+\taq\varphi \siq\tau} \left \{ - \sin\tau
\,+ \,\cos\tau \ast \right. \\
&\ast& \left. 
\sqrt{\frac{1}{\taq{\alpha_0} + \taq\varphi \left [ \taq{\alpha_0} \siq\tau -
\coq\tau \right ]}} \right \}\ .\label{eq:dbetadphi-tau}
\end{eqnarray}

\begin{figure}\begin{center}
\includegraphics[width=.6\textwidth]{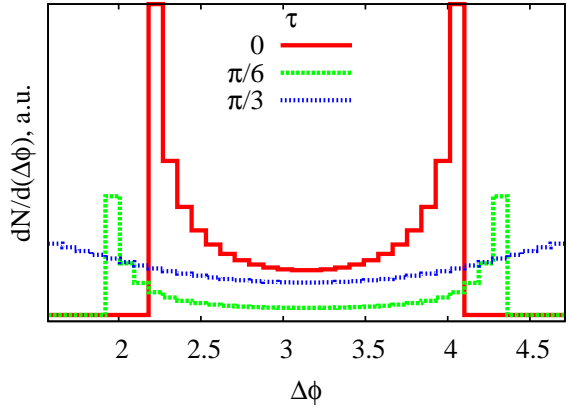}
\caption[Calculated two-particle correlations in static medium]{Calculated
two-particle correlations in static medium. Shown are the correlations for a
jet at mid-rapidity (solid line), for a jet at $\tau = 30^\circ$ (dashed
line --- $\pi/6$) and at $\tau = 60^\circ$ (dotted line --- $\pi/3$) in the
backward hemisphere. It can be seen how the maximum wanders ``outside''.}
\label{fig:two-calc}
\end{center}\end{figure}

Although equation \eqref{eq:dbetadphi-tau} seems very complicated, two
special cases can be examined very easily. For $\tau = 0$, which corresponds
to a jet in mid-rapidity, one can re-obtain equation
\eqref{eq:dbetadphi-mid}, and for $\tau = \pm \pi / 2$ one can as well
easily see that the result is $\mp 1$, which is expected because in these
cases (the jet goes along the $\hat z$-axis) $\varphi$ and $\mp \beta$ are
equal. The function is plotted for three different $\tau$, namely 0, 30 and
60 degrees, in figure~\ref{fig:two-calc}.

For certain angles $\tau$, the distribution \eqref{eq:dbetadphi-tau} is
obviously undefined. In the cases where the root gets imaginary, there are
no particles emitted, the distribution should hence be set to zero. In the
cases where the root diverges, a divergent measurement will be prevented by
the fact that what is measured in an experiment is always the average of the
distribution function over a small interval. 

However, at the border of the defined interval there is a peak. With the
modulus of $\tau$, $\abs\tau$, getting bigger, the defined interval gets
bigger. For mid-rapidity it is ${\mathcal D} = ( \pi-\acos{c_S}, \,
\pi+\acos{c_S} )$, but for all other jets it will be larger. No jet will
therefore contribute a peak between the mid-rapidity-peaks, but only
outside. Therefore, the maximum of the distribution function will shift
``outwards'', i.e.\ it will suggest a smaller speed of sound\footnote{Of
course, this could be shown by integrating equation \eqref{eq:dbetadphi-tau}
over $\tau$ and analyzing the resulting function. We rather take the
figurative approach.}.
Note that this effect always draws in one direction, and only a measurement
at absolute mid-rapidity may reveal the true speed of sound --- but then
again, this all is for a static medium only.

\chapter{MACE --- Mach Cones Evolution} \label{chap:model}

In order to study mach cones in a medium with realistic behaviour, a model
has to be built that either creates cones together with the medium (so
that the cones become an inherent part of the evolution) or one that does
add cones to the medium with hindsight, as perturbations.

The first case requires the evolution algorithm to have a much higher
spatial resolution than it would be reasonable to have for a hydro-code,
else small (and localized) perturbations would be lost very quickly.
Propagating a jet as a very strong perturbation within a locally equilibrated
medium is at least a questionable thing to do. The arising dilemma would be
eliminated if one only propagated the sound waves within the equilibrated
system and considered the jet being an external source of energy and
momentum that, if it is affected by the medium at all, only interacts
outside of the hydro framework.

In the latter case, on the other hand, backreaction towards the system has
to be neglected. This is, though, a quite reasonable choice for sound-like
perturbations which are expected to be small (else they are not sound-like
any more). But regrettably also energy-conservation is not fulfilled.
Neither the energy of the jet nor the additional momentum which in the end
will be calculated can be taken out of the system but have to be added.
Therefore, one has to assume that the total energy is much bigger than the
energy added. This, too, is not very hard to argue for; even at RHIC
energies ($\sqrt{s\ri{NN}} = 200$~GeV) adding a single $50$~GeV-Jet (which
is tremendously high) would only change the total energy by $0.13\,\%$.
\label{mark:model:energy}

Adding waves to the system after the evolution is calculated allows for a
much higher resolution; the position of a wave can in principle be specified
up to the precision of floating-point operations at the calculating machine,
whereas in the other case one was limited to the grid size used for the
system.

Our approach is to take an existing hydro-evolution and impose a jet and the
waves it creates after that as perturbations.

\section{Initialization and propagation of the jet} \label{sec:model:jet}

Only the away-side-jet is considered, since the near-side-jet is not
affected by and does not affect the medium.

The jet starts from a random position within the medium at the first
timestep calculated by the hydrodynamical code. This is, depending on the
system considered, after few $fm$ after the first collisions.

The jet is created at a given point with a probability that is proportional
to the energy-density at that point. The jet's direction is totally random.
This is a reasonable thing to demand, since we consider an equilibrated
system, in which collisions in any direction may occur. So, even with solid
angle-dependent cross-sections one will get a spherical symmetric
distribution of secondary particles\footnote{This does not apply to the
final measured momentum distribution, but it must be true if we take each
collision by itself.}. The only cut that is reasonably made is that we
exclude all jets that will not end up in the detector. More specifically, we
exclude for these studies all jets with a pseudorapidity $\abs{\eta} < 0.9$,
which is the acceptance of the ALICE-TPC \cite{else:TR_TPC_ALICE}. This
corresponds to an opening angle of $\Delta \theta \approx 88.5^\circ$ and to
a covering of about 70\ \% of the solid angle.

The jet is considered to be high energetic enough not to change its velocity
and direction. Calculation of jet quenching is not within the scope of this
thesis. It hence propagates in a straight line with speed of light through the
medium and excites sound waves as it goes along. Its (final) direction will
be used for correlation considerations in the end. When out of the medium,
the jet does not excite sound waves anymore.

\section{The waves}\label{sec:model:wave}

No premature assumptions on the shape of a resulting mach front can be made
in an unforeseeable, inhomogeneous and non-statical medium. The reshaping of
a mach front after boosting in and out of the fluid rest frame (FRF) and
considering various alignments between the jet's and the fluid's direction
have been discussed in several publications \cite{Satarov:2005mv,
Stocker:2007su, Renk:2007rv, Chaudhuri:2006qk}, but none of them took into
account a spatially inhomogeneous \cite{Satarov:2005mv, Stocker:2007su} or a
realistic, three-dimensionally evolving \cite{Renk:2007rv, Chaudhuri:2006qk}
system. Also, they considered only the wave front.

We take a different approach: we will propagate the single elementary waves
and identify the wave fronts independent of the propagation. This allows for
parts of the elementary waves to become a part of a wave front only after
some time and possible deflection. Also, some parts may be swamped away out
of the wave front. 

So, in order to model the sound waves we create a lot of logical particles,
so-called \term{wave markers}, at the position of the jet. The word
``logical'' refers to the fact that we do not assign any physical quantities
to these markers yet; they only represent the position of the wave. The
number of these wave markers per timestep is an adjustable (and numerical)
parameter, in the standard setting it is $n\ri{markers} = 1\,000$. The
entirety of the wave markers sent out at one timestep will be referred to as
one \term{elementary wave}. The markers will be assigned random directions,
so that after a short propagation in homogeneous medium one elementary wave
should indeed be a spherical wave. Such an elementary wave is created at
each timestep. Between the timesteps all ``waves'', i.e.\ all wave markers,
are propagated. 

The propagation of the wave markers is straightforward: The only assumption
made is that they move with the speed of sound relative to the fluid
wherever they are. Their direction is adjusted by relativistically adding
their initial velocity, $\vec v$, to the flow-velocity of the underlying
medium at the current point, $\vec u$.
Technically, this is done by adding the vectors corresponding to the
respective rapidities $\vec y_v$ and $\vec y_u$:
\begin{eqnarray}
\vec v\prime &=& \vec v \,\oplus\, \vec u\\
\vec y_a: && \left (
\begin{array}{rcc} 
  r &=& \frac{1}{2} \ln{\frac{1\,+\,\abs{\vec a}}{1\,-\,\abs{\vec a}}} \\
  \vartheta &=& \vartheta_a \\
  \varphi   &=& \varphi_a \end{array} \right)
\end{eqnarray}\begin{eqnarray}
\vec y\prime &=& \vec y_v + \vec y_u \qquad \mbox{(normal vector-addition)}
\label{eq:rel_add_velocity}\\
\vec v\prime: && \left ( \begin{array}{rcc} 
  \abs{\vec v\prime} &=& \frac{\exp{2 \abs{\vec y\prime}} \,-\, 1 }
             {\exp{2 \abs{\vec y\prime}} \,+\, 1 } \\
  \vartheta\prime &=& \vartheta_{y\prime} \\
  \varphi\prime   &=& \varphi_{y\prime}
\end{array} \right)
\end{eqnarray}

When a marker crosses the border of a fluid cell, then its current
propagation $\vec r\prime = \vec r \,+\, \vec v \cdot {\rm d}t$ will be
finished with the old velocity, i.e.\ no deflection happens at the border of
two cells. This is justified because the timestep ${\rm d}t$ is much smaller
than the dimensions of the cells ${\rm d}x$, ${\rm d}y$ and ${\rm d}z$, so
this will not happen all the time and will not cause too big an error.

If a marker leaves the system, it is deleted. No particle emission at this
point is assumed, and the wave is also not reflected.

\section{Freeze-out of the sound wave} \label{sec:fo_soundwave}

In the following we will discuss how to extract spectral enhancements from
the position of the wave markers. This is not at all trivial. To explain why
a certain method does or does not work, we will refer to the analytical test
case of a static medium. Unless explicitly denoted, we claim that the
argumentation is valid for non-static medium as well, but would be a lot
more complicated to explain.

\subsection[Trivial addition]{Why trivial addition does not work}
\label{ssec:fo_sw:trivial_addition}

A trivial way to evaluate the wave-markers is to assign a certain magnitude
of perturbation to each of them and add this to the current cell's momentum
and energy density. This may be done with a positive ``delta''-function that
adds a value at the exact point of the marker, or with a smoother function
that adds values at the position of the marker, but also before and after
that; presumably, since we want to consider sound-like perturbations, we
should add as much energy-momentum as we subtract at another place. The
latter might also simulate a kind of interference between different
wave-markers.

\begin{figure}\begin{center}
\includegraphics[width=.75\textwidth]{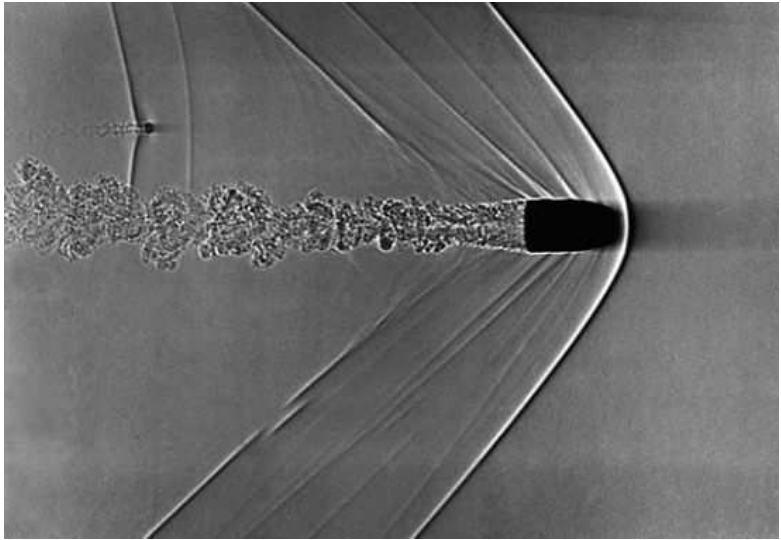}
\caption[Bullet in water]{A picture of a bullet moving through water. The
bullet is moving faster than the speed of sound in the medium, therefore a
mach cone develops. From~\cite{else:waterbullet}.}\label{fig:mach_bullet}
\end{center}\end{figure}

This way is a good physical choice in order to obtain spatial information
about the perturbations. If such a spatial picture of the mach waves is to
be made, this is the way to go. This is, in fact, what nature does, and what
helps us to make photographs of aeroplanes or bullets (see figure
\ref{fig:mach_bullet}) showing a clearly visible mach cone.

Unfortunately, a picture of the mach cone is not asked for. We try to
achieve a momentum distribution of the system, since this is what will be
measured. The momentum distribution of a system altered using the above
method will be equal to the unaltered distribution (or enhanced, if energy
and momentum have only be added). Using constant time freeze-out (see
section \ref{par:themodel}) we integrate over the whole ``picture'' and
pick up the contributions along the mach cone that we saw on the picture.
We could see it since their magnitude has been big due to the sheer number
of wave markers in close proximity. But we also pick up the contributions
opposite to it that do not contribute to the picture. They are not visible
because there are only some at relatively big distances. Nevertheless, even
if we would not discover their contribution on a spatial picture, their
integral remains the same, no matter if they are close to each other or far
apart, so they will cancel the contribution of the mach cone.

Not even the interference one might build in will help here, because
integration is a linear operation. Taking all perturbations, squaring them
and adding them to the system, which would only "count" the amplitudes and
resolve the ``integral is constant''-problem, will lose all information
about the waves direction. These effects are not unphysical and exactly what
would happen with the air perturbed by a supersonic plane. The problem
arises with the heavy-ion-way of measuring. When studying a plane we take a
picture and look at where the mach cone is. This means we look at the
amplitude. If we do not take several pictures, we have, in fact, no idea on
where the sound wave will be at a later time step. Another way on how to
watch a fast plane is to stand still and listen. We can hear the
perturbation. The analogous thing to do in the hydro-evolution is to go to
one place and wait for the wave to come --- i.e.\ freeze out at constant
place. Due to the assumption that particles will not change their direction
after freeze-out spatial information would directly be mapped to momentum
information. This method fails for the used freeze-out method (see again
section \ref{par:themodel}, page \pageref{par:themodel}).

\subsection{A coalescence-like model}\label{ssec:fo_sw:coalescence}

Following the idea that in the wave front many wave markers are close to
another and going into the same direction, it might be a good idea to only
consider markers that satisfy this condition. It would require two new
parameters; the distance in which a different marker is looked for, and a
maximal angle by which the directions of the two markers may deviate.

Unfortunately, one finds that in the analytic test-case the distance between
two markers emitted into the same direction $\alpha$ at to subsequent
time\-steps, will be $dr = dt\, \sqrt{ 1 \,+\, c_S^2 \,- \,2 c_S \cos\alpha
}$, which will be minimal at $\alpha = 0$. Forward markers can therefore not
be excluded by this method, although they do not contribute to the mach
cone.

This ansatz anyways has not much to do with interference; markers in short
distances in the direction of their movement will rather interfere than
contribute to collective movement. Markers that are next to each other,
though, will contribute. This fact might be taken into account by not using
the distance of two markers as criterion. Instead, one can model the
distance in a different way:

Usually, one can reinterpret the concept of distance between $\vec r$ and
$\vec r\prime$ as taking a sphere around the point $\vec r$ and adjusting
the radius so that $\vec r\prime$ is on the surface of that shell. The
radius $d\ri{sph}$ is then the usual --- spherical --- distance between the
points.

The idea that is to be presented here is not to take a sphere and adjust its
radius, but to take an ellipsoid with given excentricity and adjust the
remaining parameter (one of the semi axes) so that the second point is on
the shell of the ellipsoid.

For these assumptions it is reasonable to have the semi-axes of the
ellipsoid aligned along the direction $\hat v$ of the considered wave
marker; the ellipsoidal should be symmetric in the azimuthal direction with
respect to $\hat v$. In fact, the semi-axis parallel to $\hat v$ is, for the
reasons stated above, the shortest.

The (implicit) equation for an ellipsoid satisfying the above constraints
having semi-axes $A$ orthogonal and $B$ longitudinal to $\hat v$ is
\begin{equation}\label{eq:implicit_ellipse} \frac{\left (\hat v (\vec
r\prime - \vec r) \right )^2}{B^2} - \frac{\left (\hat v (\vec r\prime -
\vec r) \right )^2}{A^2} + \frac{\left (\vec r\prime - \vec r
\right)^2}{A^2} = 1 \quad .  \end{equation}

With given ratio $B = z A$ this becomes $A^2 = (1 - z^2) \left (\hat v (\vec
r\prime - \vec r) \right )^2 + \left (\vec r\prime - \vec r \right )^2$.
$A$ is what we refer to as elliptic distance $d\ri{ell}$. Note that it
becomes the spherical distance for $B \rightarrow A$, i.e.\ $\lim_{z
\rightarrow 1} d\ri{ell} = d\ri{sph}$.

In the same case as above we can now calculate $d\ri{ell}$ as a function of
$\alpha$. It turns out that we indeed have a maximum for $d\ri{ell}$ at
forward directions now instead of a minimum, and a minimum at $\cos\alpha =
\pm B^2\,c_S / (B^2\,-\,A^2)$, which is, though, always smaller than the
expected angle $\cos\alpha = c_S$.  To reach it, one would have to set $A =
0$, which is not reasonable. Using this ansatz would hence always
underestimate the angle.

\subsection[The Mantle Method]{What else can be done: The Mantle Method}
\label{ssec:fo_sw:mantle}

There \emph{is} a way of looking at a picture of a mach cone and predicting
where it will be in the next instant; we consider the movement to be
perpendicular to the wave-front. Finding out the directions perpendicular
(or parallel) to the front is easy to do by hand, one ``sees'' where the
front goes. It is not so easy to find an algorithm that discovers these
directions automatically.

Our approach is to try and reshape the region occupied by wave markers with
a set of lines, referred to as \term{wave-lines}. This surface will be
interpreted as the wave front.

The wave lines are equally distributed azimuthally around the jet's
direction $\hat v$. Their nodes are positions of marker particles. The rough
idea is to go in opposite direction of the jet, starting at its position, so
that seen from the jet it looks like standing at the pole of the earth and
considering the longitudes going away covering the surface of the earth
(apart from the earth being a sphere, not a cone).

For the construction of each wave line, a subset of all wave markers is
considered. A slice azimuthal to the jet axis is taken into account for
every wave line. The center of the azimuth, though, is not necessarily on
the trajectory of the jet, but is the geometrical center (``center of
mass'', though no masses are present) of a given elementary wave, so it
changes for markers from different elementary waves. The azimuthal slice is
classified by an angle $\phi\ri{line}$. It is the angle between the plane
spanned by $\hat v$ and the wave line on the one side and the plane spanned
by $\hat v$ and $\hat a$ on the other side, where $\hat a$ is some
arbitrary, fixed direction perpendicular to $\hat v$. We shall use the
convention
\begin{eqnarray} \label{eq:freezeoutsoundwave:a_vab} \hat a \Def \frac{\hat
v \times \hat z}{\abs{\hat v \times \hat z}}&\\ \hat b \Def \hat v \times
\hat a&,\label{eq:freezeoutsoundwave:b_vab} \end{eqnarray}
where $\hat z$ points in the z-direction\footnote{This choice for $\hat a$
is practical since we do not consider events with the jet going into the
longitudinal direction $\hat z$, so we do not have to check whether $\hat a$
is non-zero.}. We call the right-handed coordinate system $\hat v, \hat a,
\hat b$ the \index{vab-system}\term{vab}-system.

Now, we can classify the wave markers by assigning them a similarly defined
angle, $\varphi\ri{marker}$. In principle, this angle is no different from
$\phi\ri{line}$, but we allow the intersection line between the two
respective planes to change: While the $\hat v$--$\hat a$ plane remains a
constant reference, the second plane is now spanned by $\hat v$ and the
difference vector of the wave marker $\vec r_j$ and the geometrical middle
$\vec r\ri{GM}$ of its elementary wave $\vec r\ri{diff} = \vec r_j - \vec
r\ri{GM}$ (as indicated above). It should be noted that for these
considerations it is important to track the sign of the angle. 

This problem can be composed into a projection of $\vec r\ri{diff}$ onto the
$\hat a / \hat b$-plane and the angle between that projection and $\hat a$.
Considering $\vec r\ri{diff} = A \hat a\,+\,B \hat b\,+\,V \hat v$, the
projection is $\vec r\T = A \hat a \,+\,B \hat b$. The angle is then
\begin{equation} \label{eq:freezeoutsoundwave:phi_marker} \varphi\ri{marker}
= \acos{\frac{A}{\sqrt{A^2\,+\,B^2}}}\sgn{B} \end{equation}
where $A = (\vec r_j \,-\, \vec r\ri{GM}) \cdot \hat a$ and $B = (\vec r_j
\,-\, \vec r\ri{GM}) \cdot \hat b$ are the projections of $\vec r\T$ towards
$\hat a$ and $\hat b$.

For a given wave line with $\phi\ri{line}$ we define a new coordinate system
\term{vfg($\phi)$} which is defined by 
\begin{eqnarray} \hat f \Def  &\cos\phi\ri{line} \hat
a\,+\,\sin\phi\ri{line} \hat b&\text{and}\\ \hat g \Def  &\hat v \times \hat
f & .  \end{eqnarray}
Furthermore, all wave markers with $\varphi^i\ri{marker} \in \left . \left [
\phi\ri{line} - \Delta\phi / 2; \phi\ri{line} + \Delta\phi / 2 \right .
\right )$ are considered as nodes, where $\Delta\phi = 2\pi/N\ri{lines}$ is
the angle between two wave lines.

The sheer surface of the area might be very distorted or irregular due to
numerical reasons. For instance, due to the finite number of timesteps, the
surface will contain large portions of the same elementary waves (in the
limit of infinitely many elementary waves each of these only contribute one
point, which is a circle in a plane orthogonal to the jet's direction). In
order to account for this numerical problem, one has to maximize the angle
between the wave line and the jet axis when looking for the next node.  This
angle, denoted $\alpha$, between a node $\vec r_i$ and a candidate $\vec
r_j$ has to be maximized. It can be written analogous to
\eqref{eq:freezeoutsoundwave:phi_marker}:
\begin{equation}\label{eq:freezeoutsoundwave:alpha} \alpha \Def
\acos{\frac{-V}{\sqrt{V^2\,+\,F^2}}}\sgn{F} \end{equation}
where $-V$ and $F$ are the projections of $\vec r_i \,-\, \vec r_j$ onto $-
\hat v$ and $\hat f$, respectively.

\begin{figure}[t]

\begin{center}
\includegraphics[width=.5\textwidth]{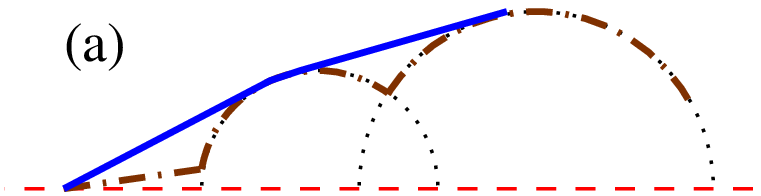}%
\includegraphics[width=.5\textwidth]{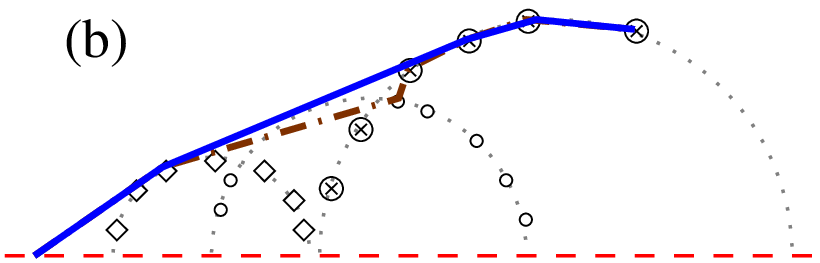}
\includegraphics[width=.50\textwidth]{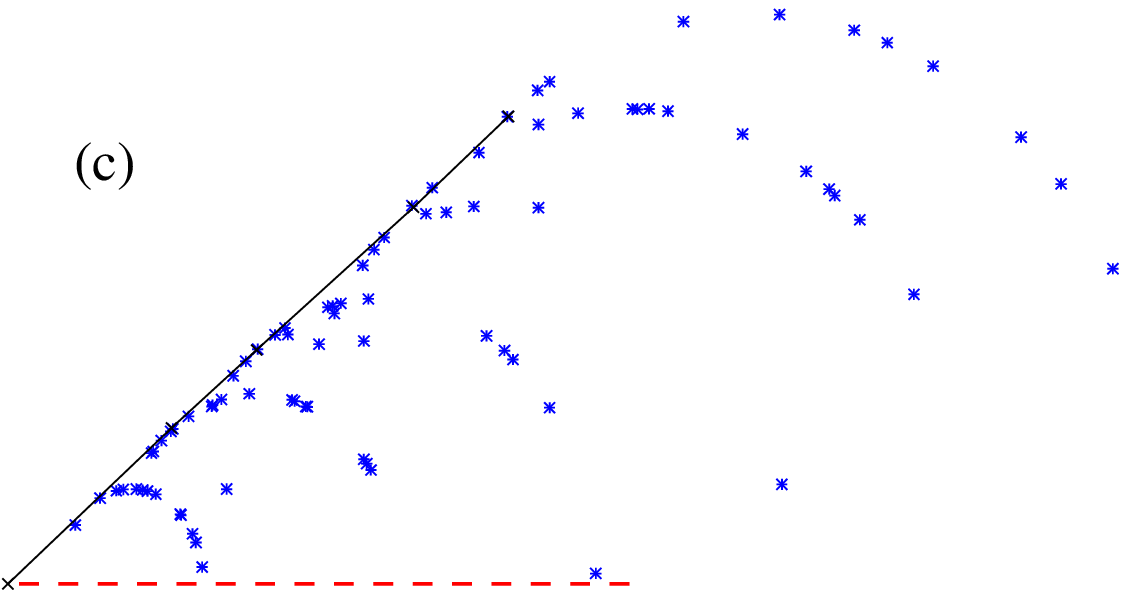}
\end{center}

\caption[Construction of the wave-lines]{A schematic view on how the lines
are constructed, assuming that all wave markers are in a common plane. In
(a) the basic smoothing is shown (maximizing the angle). In (b) the effect
of leaving out single timesteps is shown. Finally, in (c) a dynamically
created wave line from data in static medium is shown.\\ The dashed line
shows the jet-trajectory, the dotted arcs the elementary sound waves, the
dash-dotted lines the wave lines how they would be without optimization and
the solid lines the final wave lines.}
\label{fig:freezeoutsoundwave:waveline}

\end{figure}

Another numerical aspect is the random direction of the wave markers and
their finite number. It might be the case that in the area where a given
elementary wave should represent the surface there is no wave marker to
determine a possible node. Therefore, one has to give the possibility to
skip a single elementary wave and go straight onto the second next one
without any node taken from the next. Allowing for more than one elementary
wave to be ignored may conceal physical effects when collective flow is
going into different directions at different points.
Figure~\ref{fig:freezeoutsoundwave:waveline} sketches the mentioned cases
and shows the surfaces without and with the named improvements.

The search for the next node holds one additional pitfall: since the
possible wave markers are not all in one plane, a slight deviation might
occur and the wave line might finally drift out of its domain; in the worst
case, all lines would end up with the same nodes after some distance.
Therefore we have to ascertain the azimuthal position of the wave lines.
Hence, the nodes do not correspond to actual positions of wave markers, but
to their rotation onto the fixed $\hat v / \hat f$-plane. Since we can
assume that the number of lines is big enough to provide small angle
approximations for $\dd\phi$, we do not rotate, but merely project the
vector onto that plane. From a candidate $\vec r_j$ we therefore get a new
node $\vec r_{i+1}$ with
\begin{equation} \vec r_{i+1} \Def  \vec r_j - \left [ \hat g ( \vec r_j -
\vec r\ri{GM} ) \right ] \hat g \end{equation}
When looking for the next node, we have to exclude $\vec r_j$. To make the
algorithm faster one can exclude all wave markers  that lie ``behind'' the
current node (that means, closer to the position of the jet). This can be
checked by the relation $(\vec r_j - \vec r_i)(-\hat v) > 0$, from which
follows
\begin{equation} \vec r_j \hat v < \vec r_i \hat v \end{equation}

\bigskip

\begin{figure}

\begin{center}
\includegraphics[width=.5\textwidth]{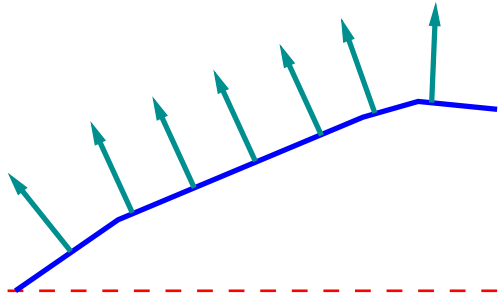}
\end{center}

\caption[Particle flow on the wave-lines]{Particle flow is assumed to be at
fixed distances (the length of the wave line is measured) perpendicular to
the current connection between next and previous position.  No additional
smoothing is done. Also it is not compensated if two or more nodes lie
between two ``freeze-out-points''.}
\label{fig:freezeoutsoundwave:particleflow}
 
\end{figure}

When all nodes are collected, particle flow is considered to be orthogonal
to the wave lines. The lines are taken \emph{as is}. To find the places on
which to insert momentum we go along the wave lines, starting from the jet,
and insert momentum after fixed distances. Momentum is inserted in the $\hat
v / \hat f$-plane orthogonal to the wave line. No special treatment for
momentum insertion at a node is done; orthogonality refers here to the part
of the wave line before the respective node (see figure
\ref{fig:freezeoutsoundwave:particleflow}). 

\section{Energy-conservation} \label{sec:model:energy_conservation}

As indicated before in the beginning of this chapter (page
\pageref{mark:model:energy}), energy cannot be conserved in this approach.
When counting the positions of momentum insertion before defining the
amplitude of insertion one can, after measuring the distance $l$ the jet has
been going already, set the total amount of energy inserted $\Delta E$ to
\begin{equation}\label{eq:energyconservation}
\Delta E = \left. \frac{\dd E}{\dd x}\right |_{E \rightarrow \infty} l \quad{,}
\end{equation}
where $\left. \dd E/\dd x \right |_{E \rightarrow \infty}$ is taken from
a different calculation. This will assure that the energy deposited in the
mach region is monotonous increasing with time and has reasonable values.

\section{Perfect Background Subtraction} \label{sec:model:background}

When the system is altered to contain the information of the sound waves,
the momentum space distribution (and in principle the whole phase space
distribution) can be evaluated to give interesting values. Actually, for
two- and three-particle correlations, background from radial, directed and
elliptic flow has to be subtracted in order to see a signal. For
experiments, background determination and subtraction is a major task.

In the theoretical approach taken, it is possible to do two different
things: One could take the original system and try to reproduce the analysis
that has to be done in experiments. This would allow for better direct
comparison to experimental results, but the data would be specific to a
special experiment and background subtraction method. Comparison to data
extracted with future, possibly better, methods would not be very easy.

The other way is what we call ``perfect background subtraction''. The idea
is to leave out the background from the beginning. In this method, the
system is not, as indicated in section \ref{ssec:fo_sw:mantle}, altered, but
the information about where the system would be altered --- the clean signal
--- is evaluated directly. This will result in the cleanest possible signal
which would represent the limit of experimental evaluations.

Here, we also have no need for correct estimation of jet energy loss, since
we only need to consider that all alterations of the system have the same
magnitude.

\chapter{Results from MACE}\label{chap:results}

There will basically be four different kinds of plots used in this chapter,
which are now explained briefly.

First of all, two-particle azimuthal correlations $\dd N / \dd(\Delta
\varphi)$ vs. $\dd(\Delta \varphi)$ will be presented, where the near-side jet
usually creates a maximum at $\Delta \varphi = 0$.  This peak will not be
visible in our plots, since in MACE there is no near-side jet. The graphs in
those plots show how many particles have been present at an angle $\varphi_i =
\varphi\ri{jet} + \Delta \varphi$. Since I have no particles in my model, the
$\hat y$-axis here is in arbitrary units, it counts insertions in the respective
direction. Cmp.\ figures \ref{fig:star_suppression} and
\ref{fig:star_twoparticle}.

A little more detailed insight to the actual situation might be taken in a
three-dimensional two-particle correlation $\dd^2 N / \dd(\Delta \varphi) /
\dd(\Delta \vartheta)$ vs. $\dd(\Delta \varphi)$ and $\dd(\Delta
\vartheta)$.  On the $\hat y$-axis, there is a second independent variable,
the difference polar angle $\Delta \vartheta = \vartheta_i -
\vartheta\ri{jet}$, analogous to $\dd(\Delta \varphi)$, which is again on
the $\hat x$-axis. The colours indicate how many particles have been at the
respective angles with respect to the jet.

Very similar is the three-dimensionally plotted  two-particle correlation
$\dd^2 N / \dd(\Delta \varphi) / \dd(\Delta \eta)$ vs. $\dd(\Delta \varphi)$
and $\dd(\Delta \eta)$, where $\dd(\Delta \eta)$ is the difference in
pseudorapidity between jet and particle.

Both these three-dimensional two-particle correlation only give new insight
for an event-by-event analysis. If data are accumulated over many events and
then correlated, they become no more meaningful as the (easier to understand and
produce) two-dimensional plots.

More insight, even over a sample of many events, can be taken with
three-particle correlations $\dd N^2 / \dd(\Delta \varphi_1) \dd(\Delta
\varphi_2)$ vs. $\dd(\Delta \varphi_1)$ and $\dd(\Delta \varphi_2)$. Here, the
colours (``$\hat z$-axis'') denote how many particles have been present at an angle
of $\Delta \varphi_1$ while another particle has been present at $\Delta
\varphi_2$. Given that the correlations are done on an event-by-event-basis, a
signal here will not be concealed or artificially created by adding the data
from many events. Cmp.\ figures \ref{fig:threeparticle_explanation} and
\ref{fig:star_threeparticle}.

The impact parameter is always given in fractions of the maximum possible
impact parameter, which is the sum of the radii of the involved nuclei.
E.g., for gold-gold collisions, $b = 10~\% \equiv 0.1 \cdot (2 \cdot
R\ri{Au}) \approx 1.36$~fm.

\section{Characterizing the algorithm}\label{sec:characterization}

The first thing that should be done with a new model or algorithm is to test
it with an analytic test case or a known approximation (whichever is
applicable), and, in the case of an algorithm, to test it against its
behaviour when changing unphysical, numerical parameters.

\begin{figure}

\includegraphics[width=.34\textwidth]{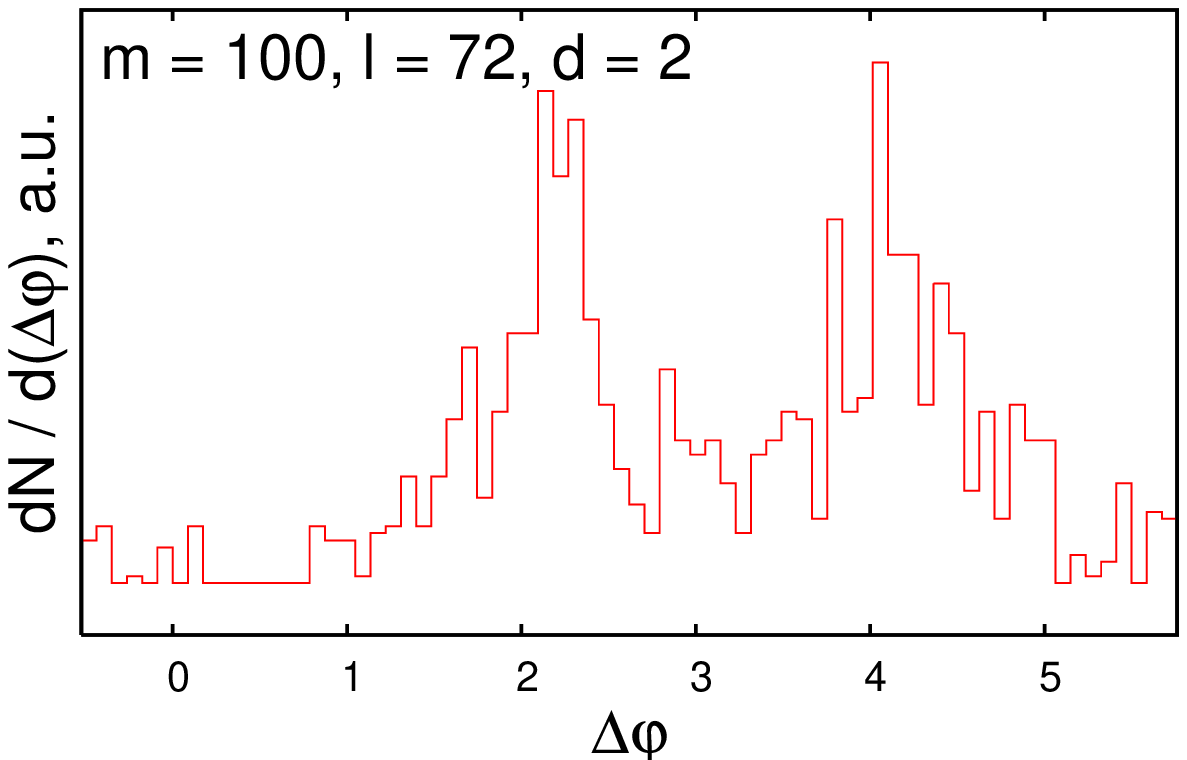}%
\includegraphics[width=.32\textwidth]{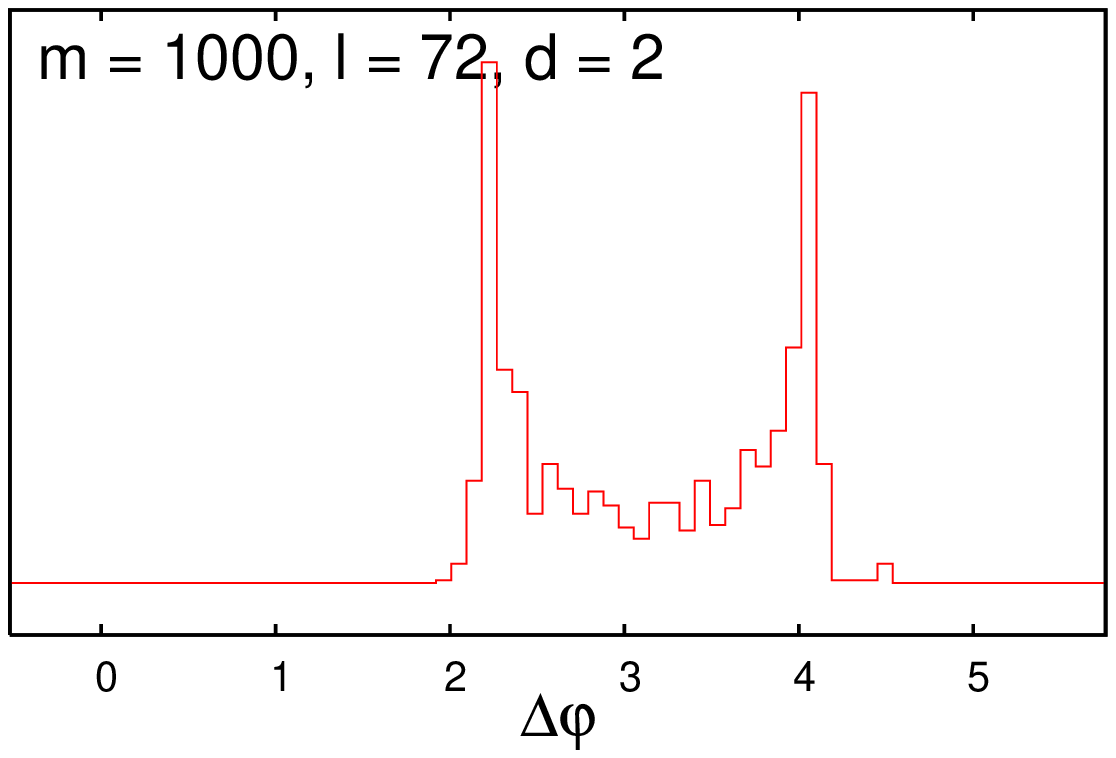}%
\includegraphics[width=.32\textwidth]{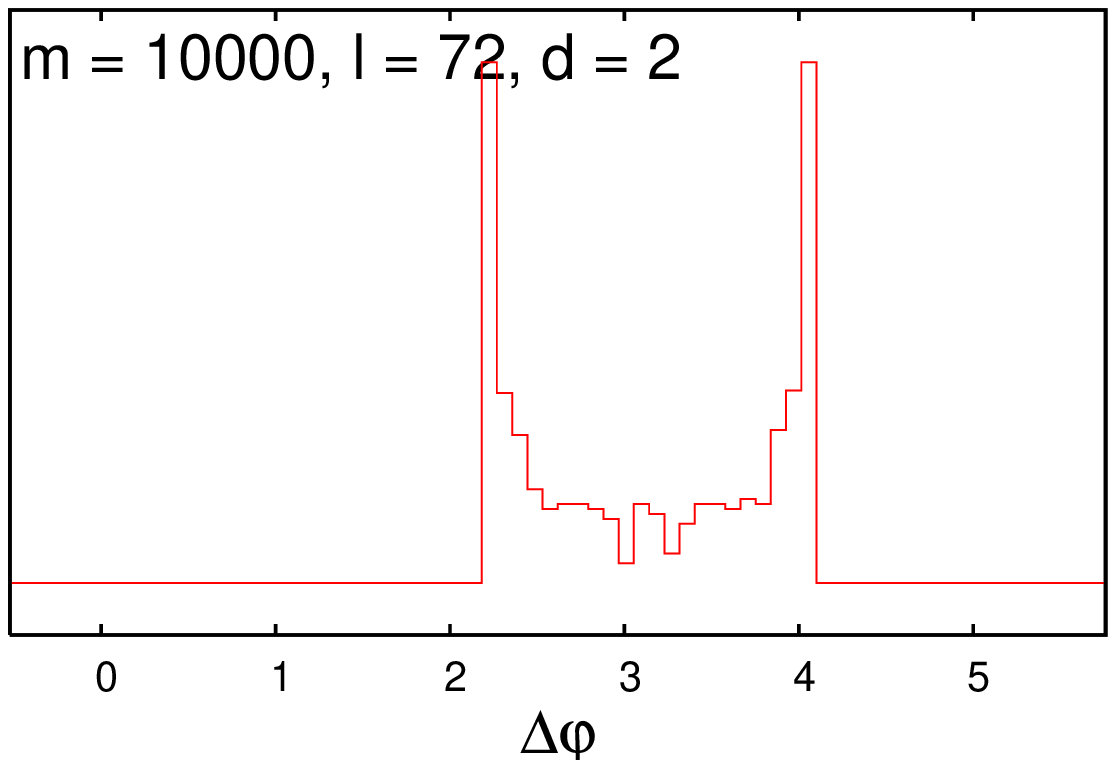}

\caption[Characterization of the code: different $m$]{The influence of the
number of wave markers $m$ on two-particle correlations (in static medium,
after 10 time steps).  From left to right, the number of wave markers goes
up by a factor of 10 between the graphs. While the raising of $m$ from 100
(left) to 1000 (middle) gives a much better signal, the improvement between
1000 and 10~000 does not change the quality a lot.} \label{fig:char:m}

\end{figure}

The analytic test case I am referring to is a static medium, i.e.\ a
system that has a collective flow velocity $\vec u \equiv 0$ everywhere and at
all timesteps. The algorithm is the one described in chapter
\ref{chap:model}, especially in sections \ref{sec:model:jet},
\ref{sec:model:wave}, \ref{ssec:fo_sw:mantle} and
\ref{sec:model:background}.

When taking the approach as described in section~\ref{sec:model:background},
namely not adding momentum to a hydro system and freezing out afterwards,
there are three numerical parameters left in the MACE-model: The number of
wave markers created at each timestep $m \equiv n\ri{marker}$, the number of
wave lines $l \equiv n\ri{lines}$ and the number of momentum insertions $d
\equiv n\ri{points}$ per one $\dd t$ length of a wave-line, where $\dd t$ is
the length of one timestep.

\begin{figure}[t]

\includegraphics[width=.34\textwidth]{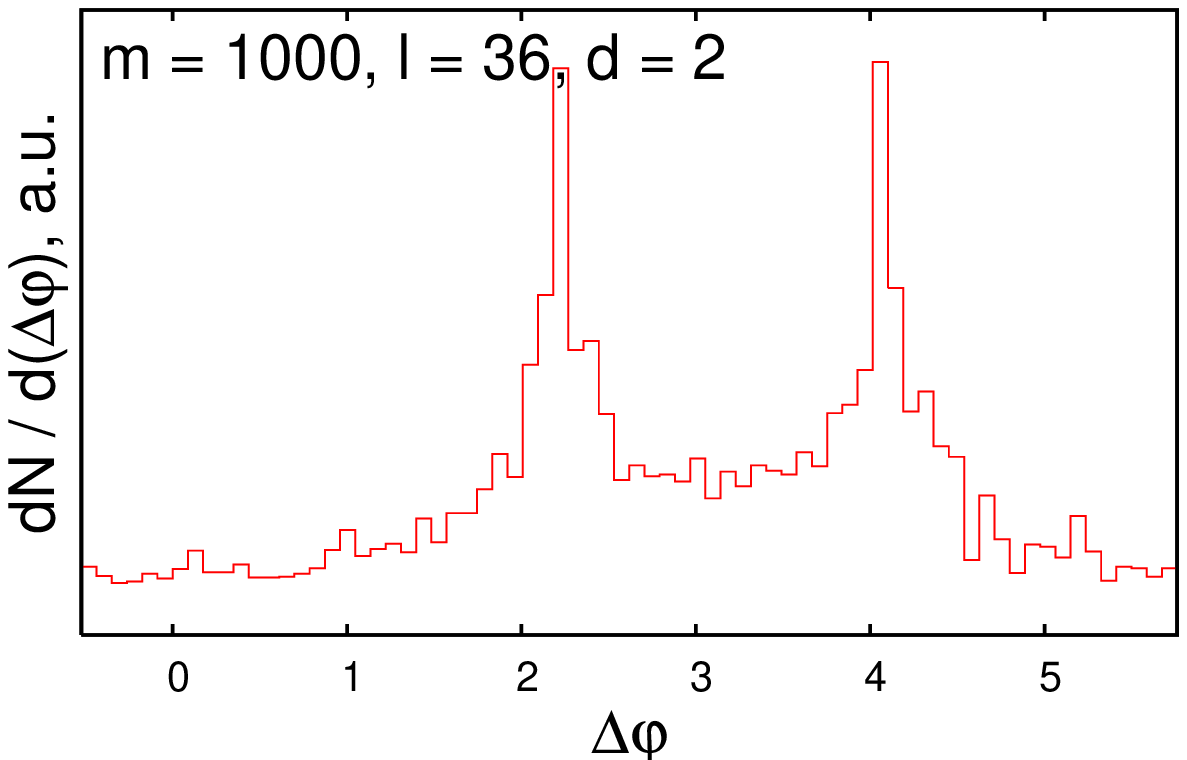}%
\includegraphics[width=.32\textwidth]{char_static_m3.eps}%
\includegraphics[width=.32\textwidth]{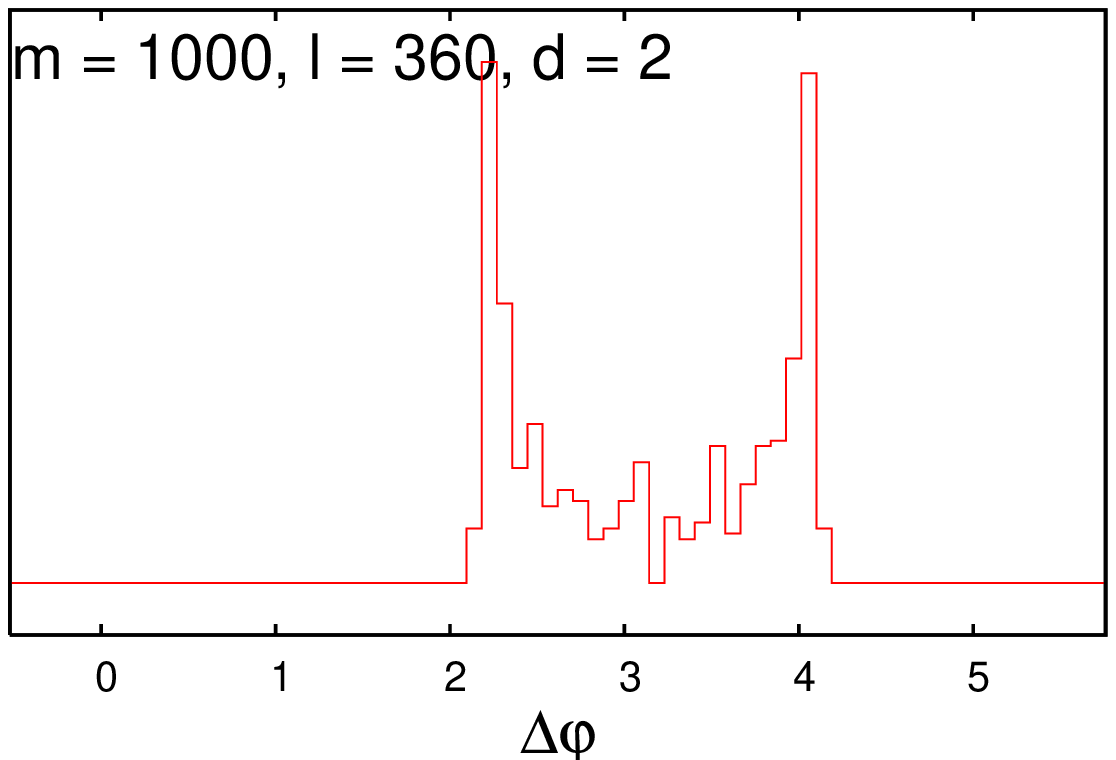}

\caption[Characterization of the code: different $l$]{The influence of the
number of wave lines $l$ on two-particle correlations (in static medium,
after 10 time steps).  From left to right, the number of wave lines goes up.
While the raising of $l$ from 36 (left, corresponds to an angle of $\Delta
\phi\ri{line} = 10^\circ$) to 72 (middle, $\Delta \phi\ri{line} = 5^\circ$)
gives a much better signal, the improvement between 72 and 360 ($\Delta
\phi\ri{line} = 1^\circ$) does not change the quality a lot.}
\label{fig:char:l}

\end{figure}

\begin{figure}[b]

\includegraphics[width=.34\textwidth]{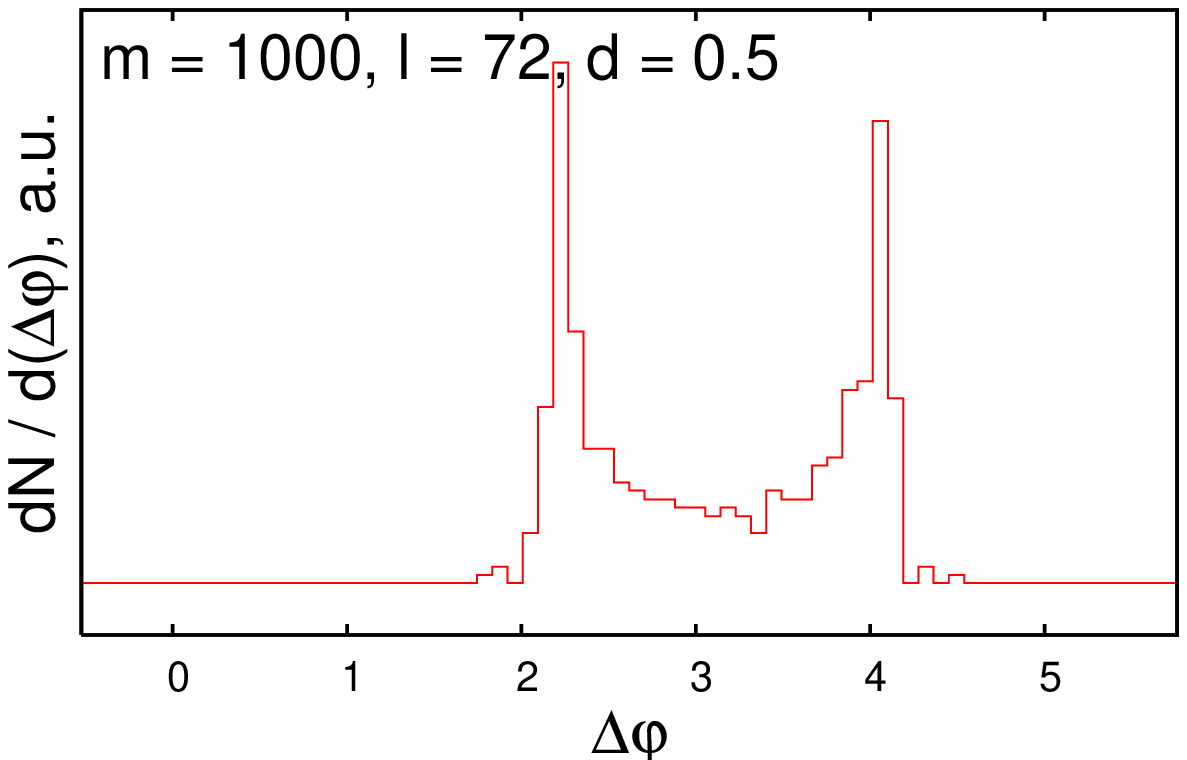}%
\includegraphics[width=.32\textwidth]{char_static_m3.eps}%
\includegraphics[width=.32\textwidth]{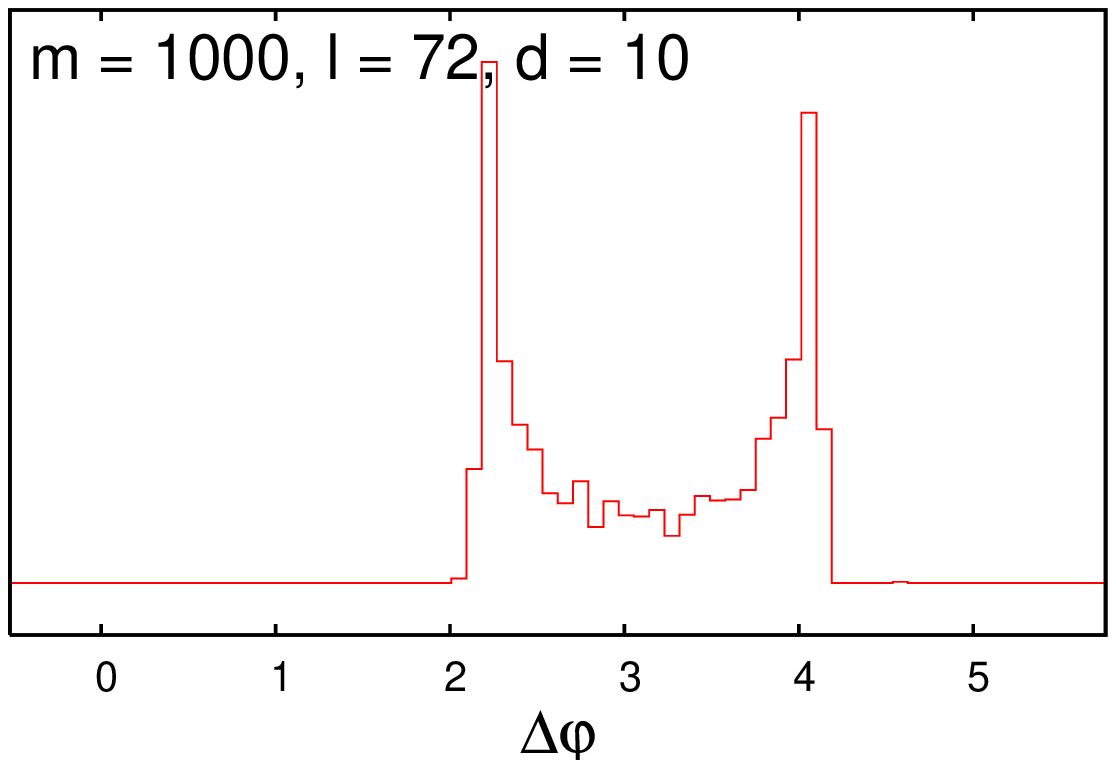}

\caption[Characterization of the code: different $d$]{The influence of the
density of momentum insertions $d$ on two-particle correlations (in static
medium, after 10 time steps).  From left to right, the density of momentum
insertions goes up. Neither step seems to change the plots a lot.}
\label{fig:char:d}

\end{figure}

With static medium and no thermal smearing being done, the resulting signal
with perfect parameters should be very sharp. With imperfectly shaped
surface, the angular distributions will be much wider, since, when the lines
are jagged, the normals will be in arbitrary directions. In the two-particle
correlations shown in figures \ref{fig:char:m}, \ref{fig:char:l} and
\ref{fig:char:d}, the expectation is not to get two delta-functions.
Instead, between the peaks the correlation function should behave similar to
an inverse circle, see section \ref{sec:analytical} and equation
\eqref{eq:dbetadphi-mid}. Outside of this region, it should indeed be zero.


Figures \ref{fig:char:m}, \ref{fig:char:l} and \ref{fig:char:d} show
two-particle correlations $\dd N / \dd(\Delta \varphi)$ for different values
of $m$, $l$ and $d$. The plot in the middle is the same in all figures; it
is made with the values that are actually used. In \ref{fig:char:m} and
\ref{fig:char:l} it can be seen that the quality improves a lot from the
worst chosen resolution to the middle one, but only minor changes appear above
that. The signal seems to be rather independent on the density of momentum
insertions $d$ (cmp.\ figure \ref{fig:char:d}).

It should be emphasized what the parameters change: A bigger $m$ makes the
wave lines smoother, while $l$ and $d$ increase the total number of momentum
insertions $N$. Correlation is an $\Order{N^2}$-algorithm, so both $l$ and $d$
should be as small as is reasonable.

All in all, it can be concluded that the algorithm shows convergent behaviour
with respect to the parameters, so finite values can be taken. In the
following subsections, unless differently denoted, all evaluations are made using
the values $m = 1\ 000$, $l = 72$ and $d = 2$.

\section{Mach cones at RHIC}\label{sec:result:RHIC}

In the following section, we discuss the results for correlation functions
in gold on gold-collisions at $\sqrt{s_{NN}} = 130$~GeV for different impact
parameters. The influence of jet triggers (both direction and origin) is
examined in section \ref{sec:result:LHC} on central LHC-collisions.

\subsection{Central collisions}\label{sec:result:RHIC:central}

\begin{figure}\begin{center}

\includegraphics[width=.75\textwidth]{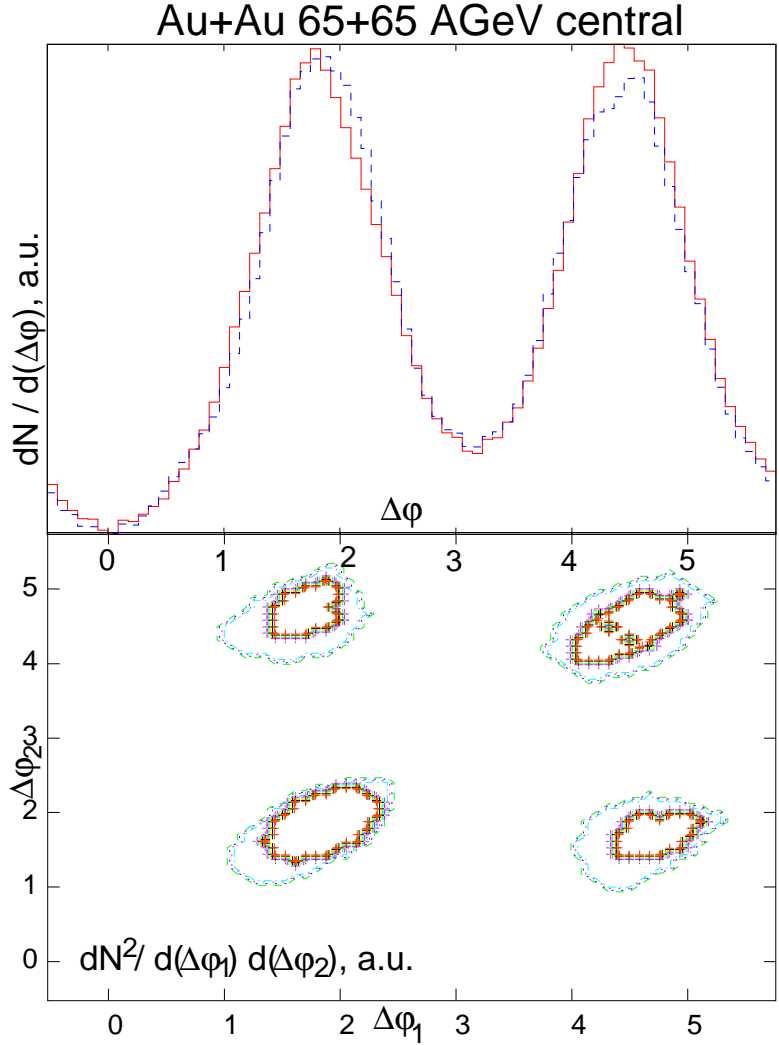}

\caption[Central RHIC 2- and 3-particle correlations]{Two- and
three-particle correlations obtained for central RHIC collisions ($b = 0~\%$,
red solid line (above) and solid lines (below), and $b = 10~\%$, blue dashed
line (above) and crosses (below)) at $\sqrt{s_{NN}} = 130$~GeV. The
dependent variable is given in arbitrary units.}\label{fig:result:rhic:c}

\end{center}\end{figure}

The central RHIC data, when averaged over 1\ 000 different jet origins and
directions, yield a double-peaked two-particle correlation function and two
off-diagonal peaks in the three-particle correlations (see
figure~\ref{fig:result:rhic:c}). The maximum of the correlation, however, is
shifted from the mach angle of $\alpha \approx 0.96$ ``outwards'' by about
$\delta \alpha \approx 0.2$~radians. The corresponding speed of sound to
this angle is $c_S^{2{\rm ,\ app.}} \approx 0.17$. This value is very close
to the expected result for a hadron-resonance gas of $c_S^{2{\rm ,\ HG}}
\approx 0.15$.

\subsection{Mid-central collisions}\label{sec:result:RHIC:mid}

\begin{figure}\begin{center}

\includegraphics[width=.75\textwidth]{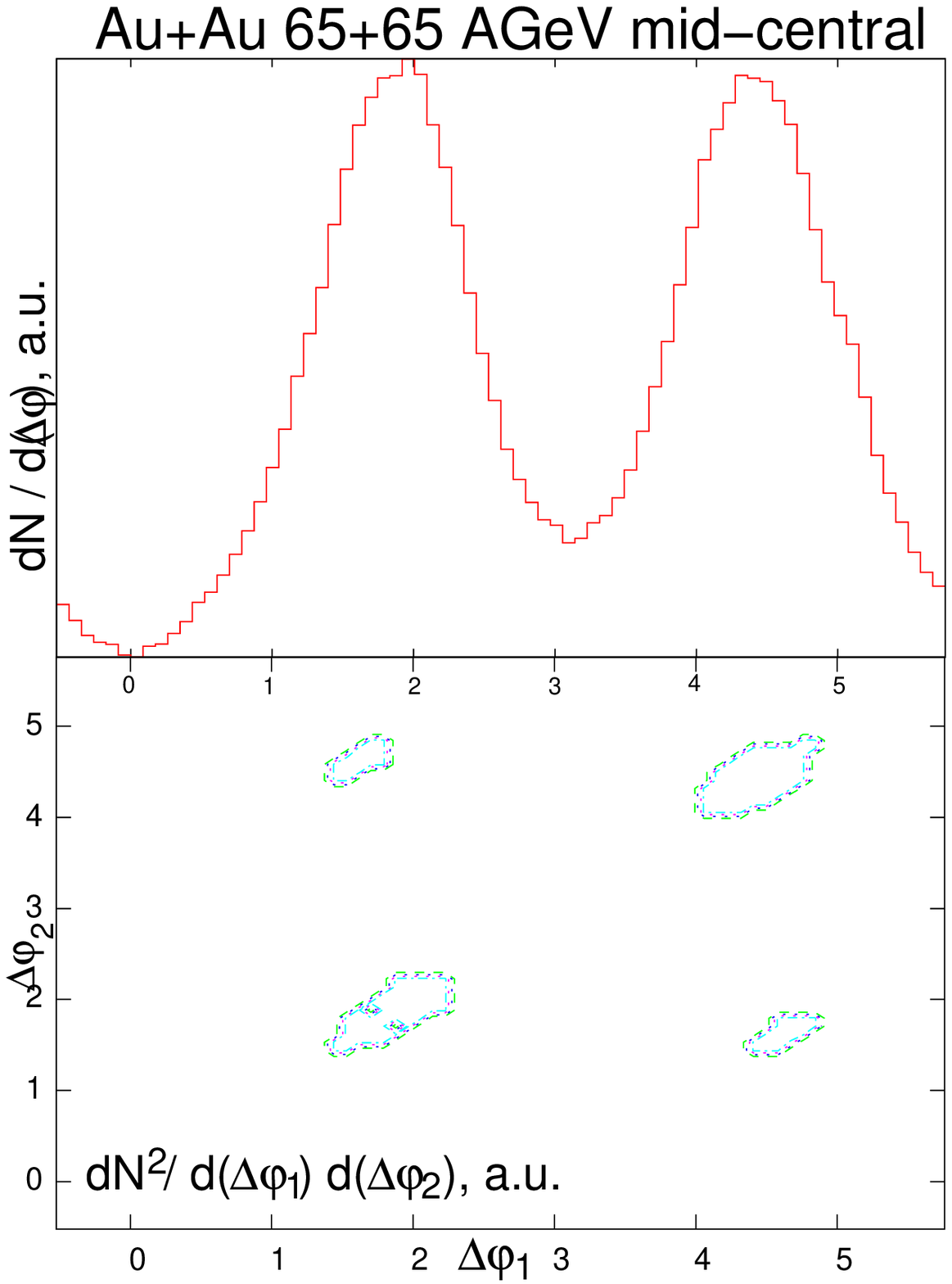}

\caption[Mid-central RHIC 2- and 3-particle correlations]{Two- and
three-particle correlations obtained for central RHIC collisions ($b =
25~\%$) at $\sqrt{s_{NN}} = 130$~GeV. The dependent variable is given in
arbitrary units.}\label{fig:result:rhic:m}

\end{center}\end{figure}

As the central RHIC data, also the mid-central RHIC data (for impact
parameter $b = 25~\%$) show two (four) peaks in the
two-(three-)particle-correlation function (see
figure~\ref{fig:result:rhic:m}). Here, the same deviation from the perfect
mach angle is found, which is again much closer to the mach angle of a
hadron-resonance gas.

\section{Mach cones at LHC}\label{sec:result:LHC}

In the following section, we discuss the results for correlation functions
in lead on lead-collisions at $\sqrt{s_{NN}} = 5.5$~TeV for central
collisions.

\begin{figure}\begin{center} %
\includegraphics[width=.75\textwidth]{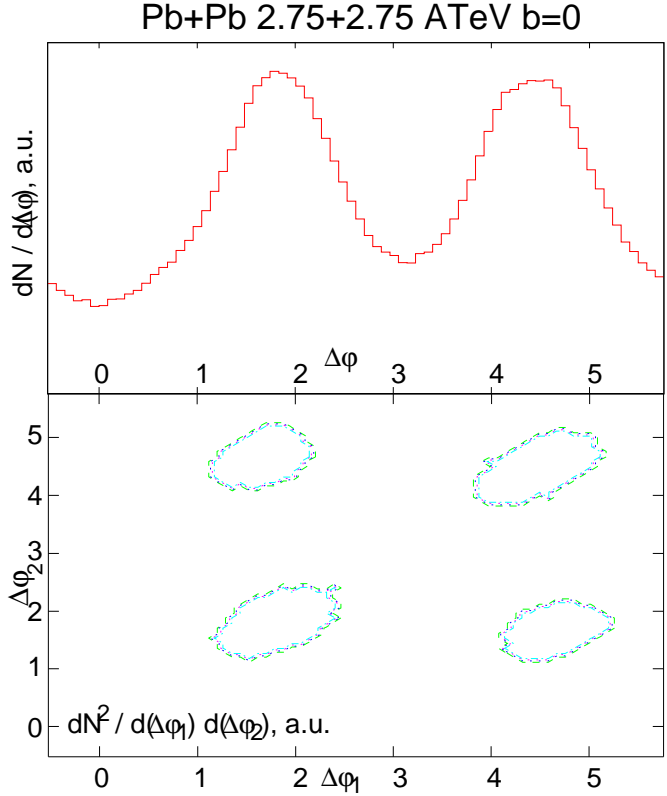} %
\caption[Correlations: central LHC; arbitrary orientations, 3D]{Two- (above)
and three- (below) particle correlations for central lead-lead-collisions at
$\sqrt{s_{NN}} = 5.5$~TeV. These figures show the sum over all correlations
from 1\ 000 events with arbitrary jet orientation and
origin.}\label{fig:lhc-mb}
\end{center}\end{figure}

\begin{figure}\begin{center} %
\includegraphics[width=.75\textwidth]{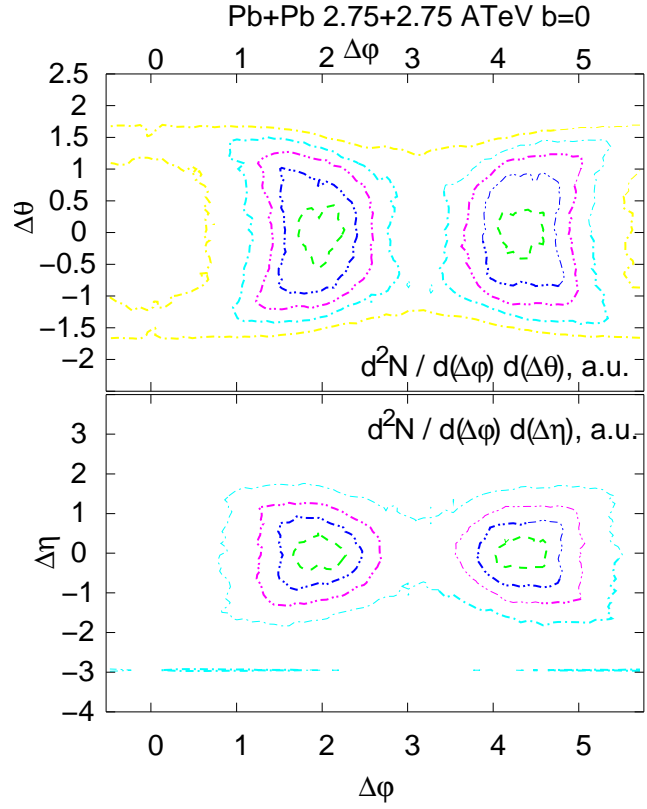} %
\caption[Correlations: central LHC; arbitrary orientations, 3D]{Two-particle
correlations for central lead-lead-collisions at $\sqrt{s_{NN}} = 5.5$~TeV
as function of $\Delta \varphi$ and $\Delta \vartheta$ (a) and $\Delta \eta$
(b). These figures show the same data as figure~\ref{fig:lhc-mb}. It can be
easily seen that these plots do not show the expected ring due to the
average over so many events.}\label{fig:lhc-mb-3d}
\end{center}\end{figure}

Correlation functions measured for LHC energies resemble the picture from
RHIC. Again, the ``minimum jet-bias''-data show a shifted maximum with
respect to the mach angle. It is also at $\Delta \phi\ri{max.} \approx
1.15$~rad, which corresponds to a speed of sound of $c_S^2 \approx 0.17 \pm
0.07$. The minimum of the correlation function is about 30~\% of the maximal
value, as has been before. Figure~\ref{fig:lhc-mb}
(lower part) proves the existence of two maxima in most events (the two
off-diagonal peaks). In figure~\ref{fig:lhc-mb-3d} 
two-particle correlations are plotted for the same set of events, with
spatial resolution now also in polar direction. This figure shows why this
way of plotting a correlation function is not very helpful; there is no ring
structure visible if one considers many events at the same time. All
information that can be obtained from these plots is that there is actually
never a signal near the near-side-jet, and finite correlations at $\Delta
\varphi \approx 0$ come from far away polar positions. The rest is contained
in the two-dimensional two-particle correlations already.

\bigskip

\begin{figure}\begin{center}
\includegraphics[width=.75\textwidth]{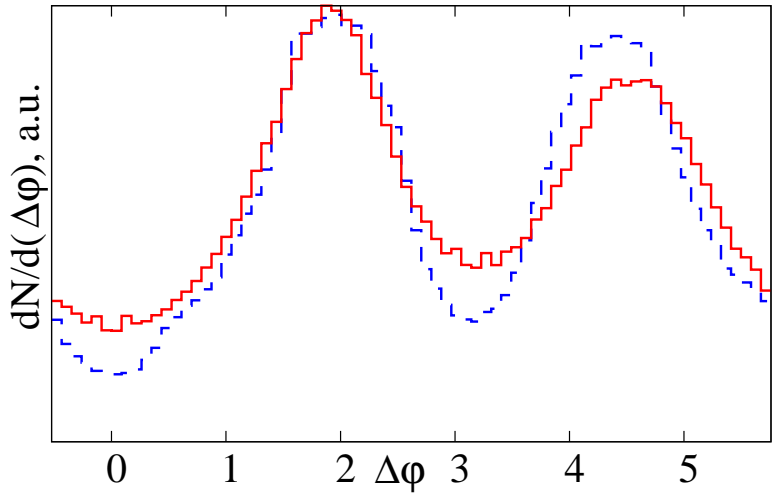}
\caption[Correlations: Rapidity trigger]{Two-particle correlations for central
lead-lead-collisions at $\sqrt{s_{NN}} = 5.5$~TeV for two different classes
of jets: The red solid line shows the result for $0.4 < \eta\ri{Jet} < 0.9$,
the blue dashed line for $\abs{\eta\ri{Jet}} <
0.1$.}\label{fig:lhc-raptrigg}
\end{center}\end{figure}

In an experiment, several triggers can be applied to the jet. While a
trigger on the azimuthal position of the jet does not make sense in the
azimuthally symmetric central collisions considered here, it is interesting
to take a look at different rapidities of the jet. In figure
\ref{fig:lhc-raptrigg}
the two-particle correlations for forward jets ($0.4 < \eta\ri{Jet} < 0.9$)
and mid-rapidity jets ($\abs{\eta\ri{Jet}} < 0.1$) are shown. This is a case
where in static medium there \emph{would} be a visible difference. Still, in
the moving matter, both correlations look rather similar. Even the
mid-rapidity data suggest a different speed of sound than is actually
present. The error bars are the same as before.

\bigskip

\begin{figure}\begin{center}
\includegraphics[width=.75\textwidth]{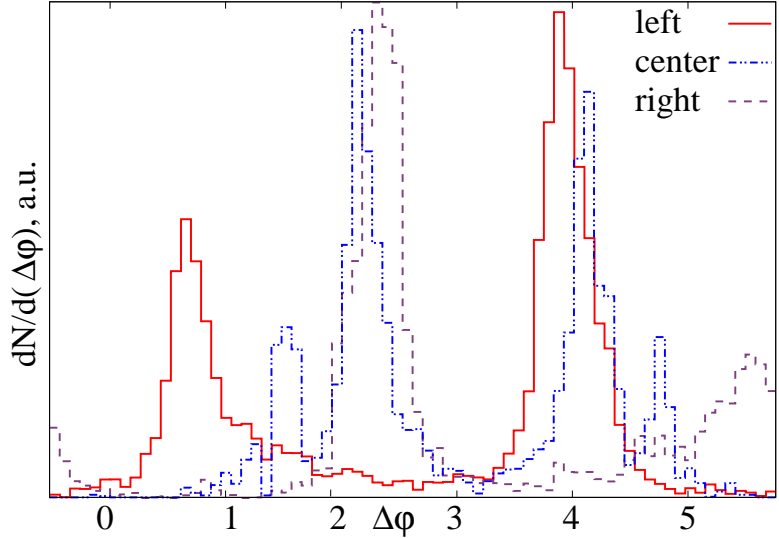}
\caption[Correlations: Different origins of the Jet]{Two-particle
correlations for mid-rapidity jets in central lead-on-lead-collisions at
$\sqrt{s_{NN}} = 5.5$~TeV.}\label{fig:lhc-lmr}
\end{center}\end{figure}

A much harder task on the experimental side is to find out where a jet came
from. Although this is not directly measurable, the effect of different
origins can be studied with MACE. The two-particle-correlation functions for
three jets going in mid-rapidity that started next to each other in the
center of the collision and left as well as right to that (at 70~\% of the
way out of the medium) are opposed in figure~\ref{fig:lhc-lmr}.
The centrally started jet, finally, shows the behaviour expected in static
medium, a symmetric correlation function (symmetric around $\Delta \varphi =
\pi$, that is), that peaks at $\Delta \varphi \approx \pi \pm 1$. It is also
revealed on this figure how much the starting point of the jet affects the
outcome of the mach cone. The left and right jet are, though, pretty
mirror-symmetric to each other, having one peak at the mach position and the
other at about $\Delta \varphi \approx \pm 0.7$.

\bigskip

Concluding the results, one can say that the double-humped structure is a
very robust signal. If there are mach cones in LHC collisions, they should
be visible. The maxima, though, are not at the naively expected angle. Only
mid-rapidity jets with symmetric correlation functions can give insight on
the speed of sound present when the cone was formed. In all other cases, the
speed of sound will be underestimated. Due to the steep behaviour of the
cosine function at the interesting values around 1 radians, even small
deviations from the perfect mach angle will give completely different
results for the speed of sound. Care should therefore be taken to read off a
speed of sound from a measured correlation function. There be dragons.

\chapter{Summary and Outlook}\label{chap:summ}

The properties of hot and dense nuclear matter have been subject to a lot of
research projects in the last decades. To study this kind of matter, atomic
nuclei (so-called \term{heavy ions}) are collided at very high center of
mass energies. From the observable particle spectra researchers try to get
information about the different properties of matter.

After the publication of some experimental results of the currently most
powerful heavy ion-accelerator, RHIC, it became probable that hydrodynamical
descriptions of nuclear matter in such collisions work pretty well.

In Hydrodynamics, the spread of sound waves is intimately connected to the
\term{Equation of State} (EoS). To get information about this it is good to
find a signal for (or \emph{from}) sound waves.

A high energetic parton travelling through the medium (a \term{jet}) might
loose its energy by bouncing off other partons. Its energy will then
propagate through the medium as sound waves. With an ultra-relativistic jet,
i.e.\ one moving practically with the speed of light, those sound waves
would --- since the speed of sound is smaller than the speed of light ---
interfere to form a cone, the so-called \term{mach cone}, as is the case
with an airplane with supersonic velocities.

In the presented thesis the model MACE (MAch Cones Evolution) is developed,
which simulates the propagation of sound waves through a medium that itself
moves with relativistic speeds in the order of magnitude of the speed of
sound.

In Chapter \ref{chap:model} it is shown how this model can be used to
simulate the sound waves of a moving source and automatically find
collective effects as a mach cone (``supersonic boom'') and analyse them.
Here, the input needed is the velocity-distribution of a medium as a
function of space and time, in order to propagate the waves with a constant
speed with respect to the local rest frame of the medium.

The elementary spherical waves are modelled by logical points. Their
entirety represents the sound wave. To recognize the mach cone
automatically, the space region in which these points are, is mantled, i.e.\
the surface is determined. That means that interference-effects are not
modelled, but a geometrical procedure is used. The surface is converted
into a signal directly, it is abdicated to first create an altered speed
profile, as it is accessible to the experiments, to re-obtain the signal
from that.

From normals to this surface a spectrum is calculated, which is measured as
increased particle emission in the experiments.

MACE is tested for its behaviour when changing numerical parameters in
section \ref{sec:characterization}. In sections \ref{sec:result:RHIC} and
\ref{sec:result:LHC} MACE is applied to the speed profiles from
gold-gold-collisions at $\sqrt{s_{NN}} = 130$~GeV (RHIC) and
lead-lead-collisions at $\sqrt{s_{NN}} = 5.5$~TeV (LHC) and the results are
presented.

It turns out that, when averaging over many jets, no significant differences
between the correlations from different starting points can be seen. Indeed,
the added-up correlation functions are indistinguishable from those expected
from static medium. Even when considering different special cases on the
direction of the jet, the picture does not change. Only when triggering on the
origin of the jet, which is experimentally the hardest of all tasks, one can
see a pronounced change in the shape of the correlations. Still, for centrally
started jets, there is no qualitative difference between LHC-data and static
medium.

In all averaged cases considered, the apparent, measured speed of sound is
smaller than what has been input to the calculation and what has been used to
propagate sound. This effect is strong enough to simulate a hadron gas with
$c_S^2 = 0.15$ in favour of the (actually present!) quark-gluon-plasma with
$c_S^2 = 1/3$. As stated above, only mid-rapidity jets starting from the
middle of the medium (recognizable by rather symmetric correlation functions)
can reveal the true speed of sound. It is emphasized that an attempt to read
of the speed of sound from an (in terms of the trigger) unqualified
correlation function is doomed to fail.

\bigskip

Since the underlying hydrodynamical algorithm (the PIC-Code, see section
\ref{par:themodel}) currently only uses an EoS with massless
Quark-Gluon-Plasma, it could not be studied how the data change with variable
speed of sound. The porting of MACE to hydro-codes that do use a more
complicated EoS and the changing of the speed of sound depending on the
properties of the underlying medium will be a future project that may give
interesting results.

The next point that may be done with MACE is comparing the results from this
approach to the data obtained by inserting a jet during the creation of the
hydro-evolution. Here, differences between those models could be studied in
detail.

\appendix

\include{conventions}

\chapter{Acknowledgements}\label{chap:acknowledgements}

My studies and this work have been graciously supported by the German
National Academic Foundation and the International Office of Frankfurt
University. I want to thank Linus for the kernel, Ian for the distribution,
Ben, Brian and Michael for minimizing the file recoveries needed (thanks for
those to Xela!), Larry for this incredibly flexible language,
Bjarne for the stiff, fast language, Jon S.\ for the best closed source I
have ever worked with, Bram for the editor and Donald and Leslie for the
typesetting system.

Thank you J\"org, Werner, Susanne, Anna-Lena, Frank, Tobias, Nathalie, and Tim
for the homely sounds, and thank you Pterry for the distraction.

\bigskip

I wish to thank everyone who supported me in the past years. First of all,
thanks to my supervisors. L\'aszl\'o P\'al Csernai has given me infrastructure
and support during the last year. He taught me a lot and I am thankful each
little bit of it. I also thank Horst St\"ocker, my official supervisor, for
supporting me and believing in me all the time since we first met. His faith
in me has always been a big reassurance in times of doubt.

Of the people I have worked with I first want to thank Marcus Bleicher who
taught me a vast amount of things about physics, programming, politics and a
lot more, and all that without me being his student. Thank you so much.

Thank you L.\ M.\ Satarov for fruitful discussions that got the last touch to
the thesis.

Also, I would like to thank my colleagues Yun Cheng, Mikl\'os Z\'et\'enyi
and P\'eter V\'an for fruitful physical discussions and Szabolcs Horvat and
Boris Wagner for their friendship. Thanks to Aasne Vik\o y for teaching me
\term{det norske spr\aa{}ket} --- it has really been a wonderful course.
Further thanks to Hilde, Maren, \O ystein, \O yvind (all of you), Hermod,
Alex, Therese, Per, Anja and Dieter.

Thank you Gabriela Meyer, clerk to Mr.\ St\"ocker, for doing infinitely many
little things that actually made it possible for me to be abroad for one year.

I want to express my gratitude to all of my friends back home in Germany for
keeping up friendship and contact over a distance of 1\ 200 kilometers and one
year, namely Sophie Kirschner, Janet Schmidt, Rainer Stiele, Heike Berleth,
Carolin B\"ose, Felix Sturm, Verena Schwenk, Luisa Ickes, Stefan Heckel, Angela
Groth, Berit K\"orbitzer and Gudrun and Gebhard Kabelitz. Thank you Christian
Stuck and my uncle Klaus Voigtmann for saving me from the downs of evil Lord
Bill and teaching me the (regrettably) Secret Ways Of The Penguin. Especially
the shared evenings on the riverside of the Main with Stefan Salm have been a
good incentive to finish quickly and come back, and I want further to thank my
best friend Johannes Schwenk for his support at the right times.

Thanks to the whole of the students council (Fachschaft Physik) and the guitar
orchestra of the music school for youngsters (Jugendmusikschule) Frankfurt for
standing me and my moods for so long time and giving me so many hours of
pleasure throughout the years.

The deepest gratitudes are expressed last: my parents Marieluise and Peter
have supported me all my life and taught me how to live. My Ultra Krasse
Sister UKS Melanie has shown me a lot beautiful sides of life and shares
many nice experiences with me.

Finally, there is only one more person to mention. My girlfriend Hannah
Petersen, who has been and still is dear friend, support, partner, ray of hope,
advice, encouragement, reason to live, reason to hurry, reason to be good,
dream, corrector and everything else I needed, and most of the time all this at
once, to me, has been the best help I could imagine. Without her, I would not
have finished this thesis.

\bigskip

Oh, by the way --- thank you to \emph{who-ever} invented \"Abbelwoi.

\chapter{Legal stuff}

\section[Zusammenfassung (German abstract)]{Deutsche Zusammenfassung der
Diplomarbeit --- German Abstract to the thesis}

Die Eigenschaften von hei\ss{}er, dichter Kernmaterie sind Gegenstand vieler
For\-schungs\-vor\-ha\-ben der letzten Jahrzehnte. Um solche Materie zu
untersuchen, werden Atomkerne (sog.\ \term{Schwerionen}) bei extrem hohen
Schwerpunktsenergien zur Kollision gebracht. Aus den dabei messbaren
Teilchenspektren versuchen die Forscher, Aufschl\"usse \"uber verschiedenste
Eigenschaf\-ten der Materie zu erlangen.

Nach Ver\"offentlichung einiger Messergebnisse des momentan
leis\-tungs\-f\"ah\-igsten Beschleunigers f\"ur schwere Atomkerne, RHIC,
scheint es wahrscheinlich, dass hydrodynamische Beschreibungen f\"ur
Kernmaterie in solchen Kollisionen recht gut sind.

In der Hydrodynamik ist die Ausbreitung von Schall eng mit der sog.\
\term{Zustandsgleichung} der Materie verbunden. Um Aufschluss \"uber diese
zu erhalten, ist es daher gut, ein Signal f\"ur bzw.\ von Schallwellen zu
finden.

Ein sich durch das Medium bewegendes hochenergetisches Parton (ein
\term{Jet}) kann eventuell durch St\"o\ss{}e mit anderen Partonen Energie
verlieren, die dann in Form von Schallwellen durch das Medium propagiert
wird. Bei einem ultrarelativistischen Jet, also einem, der sich praktisch
mit Lichtgeschwindigkeit bewegt, w\"urden sich --- da die
Schallgeschwindigkeit kleiner als die Lichtgeschwindigkeit ist --- solche
Schallwellen zu Kegeln (sog.\ \term{Mach\-ke\-gel}) \"uberlagern, wie bei
Flugzeugen mit Geschwindigkeiten jenseits der Schallmauer.

In der vorliegenden Arbeit wird das Modell MACE (MAch Cones Evolution ---
Mach-Kegel-Entwicklung) entwickelt, das die Propagation von Schallwellen
durch ein Medium simuliert, das sich selbst mit relativistischen
Ge\-schwin\-dig\-kei\-ten in der Gr\"o\ss{}enordnung der
Schallgeschwindigkeit bewegt.

In Kapitel \ref{chap:model} wird gezeigt, wie dieses Modell benutzt werden
kann, um die Schallwellen einer sich bewegenden Quelle zu simulieren und
kollektive Effekte wie einen Machkegel (``\"Uberschallkegel'') automatisch
zu finden und zu analysieren. Hierbei wird als Input die
Geschwindigkeitsverteilung eines Mediums als Funktion von Ort und Zeit
benutzt, um die Wellen mit konstanter Geschwindigkeit im Ruhesystem des
Mediums zu propagieren. 

Die elementaren Kugelwellen werden dabei durch logische Punkte modelliert,
deren Gesamtheit die Schallwelle darstellen. Um den Machkegel automatisch zu
erkennen, wird die Raumregion, in der sich diese Punkte aufhalten,
ummantelt, d.h.\ die Oberfl\"ache wird bestimmt. Das bedeutet, dass nicht
Interferenz-Effekte modelliert werden, sondern geometrische \"Uberlegungen
benutzt werden. Die Oberfl\"ache wird direkt in ein Spektrum umgewandelt, es
wird darauf verzichtet, zun\"achst ein ver\"andertes Geschwindigkeitsprofil
zu erstellen, wie es den Experimenten direkt zug\"anglich ist, um dann das
Signal daraus zu ermitteln.

Aus Normalen auf dieser Oberfl\"ache wird dann ein Spektrum ermittelt, das
im Experiment als erh\"ohte Teilchen-Emission gemessen wird.

MACE wird in Kapitel \ref{chap:results} zun\"achst auf sein Verhalten
gegen\"uber der Ver\"anderung von numerischen Parameter getestet (Abschnitt
\ref{sec:characterization}). In den Abschnitten \ref{sec:result:RHIC} und
\ref{sec:result:LHC} wird MACE dann auf die Geschwindigkeitsprofile von
Gold-Gold-Kollisionen bei $\sqrt{s_{NN}} = 130$~GeV (RHIC) und
Blei-Blei-Kollisionen bei $\sqrt{s_{NN}} = 5.5$~TeV (LHC) angewandt und
Ergebnisse werden vorgestellt.

Es zeigt sich, dass, wenn \"uber genug Jets gemittelt wird, keine
signifikanten Unterschiede zwischen den Korrelationen von verschiedenen
Startbedingungen sichtbar sind. In der Tat sind die (aufaddierten)
Korrelationsfunktionen nicht zu unterscheiden von denen, die bei statischem
Medium erwartet werden. Sogar, wenn man verschiedene Spezialf\"alle f\"ur die
Richtung des Jets betrachtet, \"andert sich das Bild nicht. Nur bei triggern
auf den Ursprung des Jets, was eine sehr schwierige experimentelle Aufgabe
ist, kann man eine ausgepr\"agte Ver\"anderung in der Form der Korrelationen
sehen. Doch wieder ergibt sich f\"ur zentral gestartete Jets keine qualitative
Ver\"anderung zwischen LHC-Daten und statischem Medium.

In allen betrachteten, gemittelten F\"allen verschieben sich die
Korrelationsfunktionen so, dass man naiv eine Schallgeschwindigkeit messen
w\"urde, die um etwa das eineinhalbfache kleiner ist als die tats\"achlich
benutzte Schall\-ge\-schwin\-dig\-keit. Dadurch kann die Existens einer anderen
Form von Materie (Hadronengas statt Quark-Gluonen-Plasma) vorget\"auscht
werden.  Wie oben angedeutet, k\"onnen nur Jets in Mittrapidit\"at, die in der
Mitte des Mediums starten (diese Jets sind erkennbar an einer symmetrischen
Korrelationsfunktion), die wahre Schallgeschwindigkeit aufzeigen. Es wird
unterstrichen, dass jeglicher Versuch, eine Schallgeschwindigkeit an einer
Korrelationsfunktion abzulesen, zum Scheitern verurteilt ist, wenn man nicht
genau wei\ss{}, auf was getriggert wurde oder der Trigger nicht auf das
Problem passte.

\cleardoublepage

\section[Erkl\"arung zur Diplomarbeit (Declaration)]{Erkl\"arung zur
Diplomarbeit}

Ich versichere hiermit, dass ich die vorliegende Arbeit selbst\"andig
verfasst, keine anderen als die angegebenen Hilfsmittel verwendet und
s\"amtliche Stel\-len, die benutzten Werken im Wortlaut oder dem Sinne nach
entnommen sind, mit Quellen- bzw. Herkunftsangaben kenntlich gemacht habe.

\bigskip

Frankfurt am Main, den 11. Juni 2007

\vspace{5cm}

{\tiny Bj\o{}rn B\"auchle}

\end{document}